\documentclass[iop]{emulateapj}
\usepackage{apjfonts} % Times fonts

\usepackage{graphicx}

\usepackage{amsmath,amssymb,graphicx,twoopt,lineno,color,soul}
\usepackage{natbib}
\usepackage{threeparttable}
\usepackage[maxfloats=50]{morefloats}

\def\nms{\mathsurround=0pt}

\def\oversim#1#2{\lower 2pt\vbox{\baselineskip 0pt \lineskip 1pt \ialign{$\nms#1\hfil##\hfil$\crcr#2\crcr\sim\crcr}}} 

\shorttitle{Rotation Reversals in OJ\,287}
\shortauthors{Cohen et al.}

\makeindex
\citeindextrue

\received{January 30, 2018}
\accepted{June 5, 2018}
\journalinfo{Astrophysical~journal}
\submitted{}
\begin{document}
\def\deg{\ifmmode^\circ\else$^\circ$\fi}

\title{Reversals in the Direction of Polarization Rotation in OJ\,287}

%\correspondingauthor{M.~H. Cohen}
%\email{mhc@astro.caltech.edu}

\author{M.~H.~Cohen\altaffilmark{1},
H.~D.~Aller\altaffilmark{2},
M.~F.~Aller\altaffilmark{2},
T.~Hovatta\altaffilmark{3},
P.~Kharb\altaffilmark{4},
Y.~Y.~Kovalev\altaffilmark{5,6,7},
M.~L.~Lister\altaffilmark{8},
D.~L.~Meier\altaffilmark{9},
A.~B.~Pushkarev\altaffilmark{10,5},
\and
T.~Savolainen\altaffilmark{11,12,7}
}

\altaffiltext{1}{Department of Astronomy, California Institute of Technology, Pasadena, CA 91125, USA; mhc@astro.caltech.edu}

\altaffiltext{2}{Department of Astronomy, University of Michigan, 311 West Hall, 1085 S. University Avenue, Ann Arbor, MI 48109, USA}

\altaffiltext{3}{Tuorla Observatory, University of Turku, V\"ais\"al\"antie 20, 21500 Piikki\"o, Finland}

\altaffiltext{4}{National Centre for Radio Astrophysics (NCRA-TIFR) S. P. Pune University Campus, Post Bag 3, Ganeshkhind Pune 411 007, India}

\altaffiltext{5}{Astro Space Center of Lebedev Physical Institute, Profsoyuznaya 84/32, 117997 Moscow, Russia}

\altaffiltext{6}{Moscow Institute of Physics and Technology, Dolgoprudny, Institutsky per. 9, Moscow region, 141700, Russia}

\altaffiltext{7}{Max-Planck-Institut f\"ur Radioastronomie, Auf dem H\"ugel 69, 53121 Bonn, Germany}

\altaffiltext{8}{Department of Physics and Astronomy, Purdue University, 525 Northwestern Avenue, West Lafayette, IN 47907, USA}

\altaffiltext{9}{Department of Astronomy, California Institute of Technology, Pasadena, CA 91125, USA}

\altaffiltext{10}{Crimean Astrophysical Observatory, Nauchny 298409, Crimea, Russia}

\altaffiltext{11}{Aalto University Mets\"ahovi Radio Observatory, Mets\"ahovintie 114, 02540 Kylm\"al\"a, Finland}

\altaffiltext{12}{Aalto University Department of Electronics and Nanoengineering, PL 15500, FI-00076 Aalto, Finland}

                   %%%%   ABSTRACT   %%%%

\begin{abstract}

We have obtained a smooth time series for the Electric Vector Position
Angle (EVPA) of the blazar OJ\,287 at centimeter wavelengths, by
making $\pm n\pi$ adjustments to archival values from 1974
to 2016.  The data display rotation reversals in which the EVPA rotates
counter-clockwise (CCW) for $\sim 180\deg$ and then rotates clockwise
(CW) by a similar amount. The time scale of the rotations is a few weeks
to a year, and the scale for a double rotation, including the reversal,
is one to three years.  We have seen four of these events in 40 years. A
model consisting of two successive outbursts in polarized flux density,
with EVPAs counter-rotating, superposed on a steady polarized jet, can
explain many of the details of the observations. Polarization images
support this interpretation.  The model can also help to explain similar
events seen at optical wavelengths. The outbursts needed for the model
can be generated by the super--magnetosonic jet model of \citet{NGM10}
and \citet{NM14}, which requires a strong helical magnetic field. This
model produces forward and reverse pairs of fast and slow MHD waves, and
the plasma inside the two fast/slow pairs rotates around the jet axis,
but in opposite directions.

\end{abstract}


                 %%%  INTRODUCTION  %%%

\section{Introduction}
\label{s:intro}

Many active galactic nuclei (AGN) show a one-sided jet that can be traced
inward to a few pc from the massive black hole that powers the system. This
one-sidedness is a relativistic effect, in which radiation from the
jet, which is composed of plasma flowing relativistically, is strongly
boosted when the observer is near the axis, while the counter-jet is
strongly de-boosted. The jet may also contain bright features that move
superluminally downstream; i.e., their apparent velocity in the plane
of the sky is greater than $c$, the speed of light. In these cases the
observed time scale is shrunk, as the emission region follows closely
behind its own radiation. This reduced time scale is also partly
responsible for the rapid variability that is seen in many AGN.

Many of these jets are highly polarized, and both the fractional
linear polarization and the electric vector position angle (EVPA) can be
variable. The EVPA is measured North through East on the sky, and
its variation will be our main concern in this
paper. In BL Lac the EVPA tends to point along the jet \citep{OG09},
and this means that in the jet the transverse component of the magnetic
field is dominant. The EVPA can point along the jet even around a bend
\citep{OG09}, and this is taken as a sign that the transverse field
is toroidal and that the field configuration is generally helical
\citep{C15}. The jet appears to be a magnetic structure that can
support MHD waves. BL Lac has been analyzed with this assumption; the
superluminal components were taken as fast or slow magnetosonic waves,
and the downstream propagation of the bent structure could be regarded
as an Alfv\'en wave \citep{C14,C15}.

A gradient of the Faraday Rotation Measure (RM) across the jet, especially
if there is a sign reversal across the jet, is another indication of
toroidal magnetic fields, since the RM is proportional to the component of
magnetic field along the line-of-sight. In a recent paper \citet{GNR18}
provide a list of 52 AGN that have reliable detections of transverse RM
gradients, and 5 of these show time variability.

In this paper we are concerned with one particular AGN, the BL
Lacertae object OJ\,287, which is highly active at all wavelengths.
We have made images of its jet with the VLBA\footnote{The Long Baseline
Observatory and the National Radio Astronomy Observatory are facilities
of the National Science Foundation operated under cooperative agreement
by Associated Universities, Inc.}, a high-resolution radio instrument
with EW resolution $\rm \sim 0.6~milliarcsec~(mas)$ at $\lambda
\approx~2$cm. OJ\,287 has redshift $z=0.306$ giving a linear scale of
4.48~pc\,mas$^{-1}$; thus we can probe OJ\,287 at scales of about one pc.
OJ\,287 is not in the RM gradient list of \citet{GNR18}, but \citet{MG17}
have tentatively identified it as having a transverse RM gradient.

OJ\,287 has provided another reason to think that the jets of AGN
are threaded by helical magnetic fields. \cite{C17} has studied the
evolution of the ridge lines of OJ\,287, and has shown that they are
twisted, and can be interpreted as sections of a rotating helix.
%% NEW SENTENCE
In the present paper the model we use contains a rotating helix,
and the observations show that it has positive (right-hand) helicity.

%% OLD and deleted
%The helix probably
%has positive (right-hand) helicity, and this also is the sense we find
%in the present paper, when using our model for the EVPA rotations.

At optical wavelengths OJ\,287 shows flares, roughly 12 years apart,
whose timimg can be fit to a model consisting of a binary black hole
system, including spin and gravitational radiation in addition to
the orbital parameters.  This model has successfully predicted the
appearance of flares in 2006--2010 and in 2015 \citep{Val11, Val16}.
In terms of kpc-scale radio morphology and power, OJ\,287 exhibits
both Fanaroff-Riley Type I and Type II characteristics; i.e.,
FR-I morphology and FR-II radio power. It  is an exception to the
simple Unified Scheme which proposes that BL Lac objects are pole-on
counterparts of FRI radio galaxies \citep{K15,S15}.

In this paper we concentrate on the EVPA of OJ\,287 at radio wavelengths,
and report the observation of $rotation~reversals$.  One of these
consists of a large counter-clockwise (CCW) swing in EVPA, followed
closely by a similar but clockwise (CW) swing.  Variations in the EVPA
of AGN, including OJ\,287, have a long history of study.  \citet{HB84}
measured the optical polarization of OJ\,287 over a 4-day period and found
rotations in time and also variations in frequency.  \citet{RGW87}
made early VLBI observations of OJ\,287 that separated the core and
the jet components, and showed that their polarizations changed over a
one-year interval. \citet{KIM88} observed a steady swing of 80\deg~in
the EVPA in 5 days at radio wavelengths, and a nearly-simultaneous swing
of 120\deg~in 7 days at optical wavelengths. A close correlation of
radio and optical EVPA rotations has also been reported by \citet{GRS06}
and by \citet{DA09}.

\citet{Vil10} have made extensive optical observations of the
EVPA of OJ\,287.  They showed that the EVPA has a long-term preferred
value, 170\deg, although it often appears to be chaotic. Currently,
the RoboPol program \citep{BP15, BP16} is making optical polarization
measurements for many AGN, including OJ\,287.

This paper is organized as follows.  In Section~\ref{s:obs}
we discuss the observations, and first show the EVPA data from
the archives. These data are erratic in some time periods, and
in Figure~\ref{evpa7416}a we smooth the EVPA by adding $\pm n\pi$ as
appropriate. This smoothing allows us to see four rotation reversals.
In Section~\ref{s:spurious} we briefly consider the possibility
that the rotations and reversals are spurious, and conclude that
they are not.  The reversals themselves are described in detail in
Section~\ref{s:reversals}.  In Section~\ref{s:stokes_vectors} we propose
a two-component model to explain an EVPA rotation as a flux density
outburst with a rotating EVPA, superposed on a steady jet component.
Two of these outbursts in succession, with counter-rotating EVPAs,
generate the reversal.  We describe a simple geometry with a relativistic
jet containing a helical magnetic field that can make counter-rotating
bursts in Section~\ref{s:geometry}, and in Section~\ref{s:super} suggest
that the super-magnetosonic jet model of \citet{NGM10} and \citet{NM14}
can help to explain the observations.

Section~\ref{s:optical} briefly describes some aspects of the optical
observations of OJ\,287.  Section~\ref{s:discussion}, the Discussion,
comments on the time scales for the rotation reversals, the many outbursts
without an EVPA rotation, and on how our reversals contain 12-year
separations, the same separation that is found for the repeating optical
flares.  Section~\ref{s:conclusions} contains a Summary and Conclusions.

\section{Observations}
\label{s:obs}

OJ\,287 is a rapidly varying source, and at radio wavelengths the EVPA
can change on a time-scale of days. On the other hand, the rapid EVPA
changes occur episodically and are unpredictable; so that, to capture
the full story of the EVPA, observations need to be made every few
days and the series has to last for many years.  The archives of the
University of Michigan Radio Astronomy Observatory (UMRAO) provide data
that meet this need \citep{AA85}.  They comprise measurements of flux
density ($F$) and polarized flux density that were made
every few days with a 26--m dish, and span the years 1975-2012. However,
OJ\,287 passes close to the Sun every year, and 1 or 2-month gaps in the
data do occur regularly, as seen in the graphs below.  Only points with
$\sigma (\mathrm{EVPA})< 14.3\deg$ are used here; this is equivalent to
limiting the signal-to-noise ratio of the linearly-polarized flux density,
$P\times F$ (hereafter simply called $PF$) to $\mathrm{SNR}(PF)>2$. 
Here $P$ is the fractional linear polarization.  Each
UMRAO point is a one-day average.

%%%%%%%%%%%%%%%%%%%%%%%%%%%%%%%%%%%%%%%%%

%% FIGURE 1
\begin{figure*}
\centering
\includegraphics[width=\textwidth,trim=0cm 12cm 1cm 0.5cm]{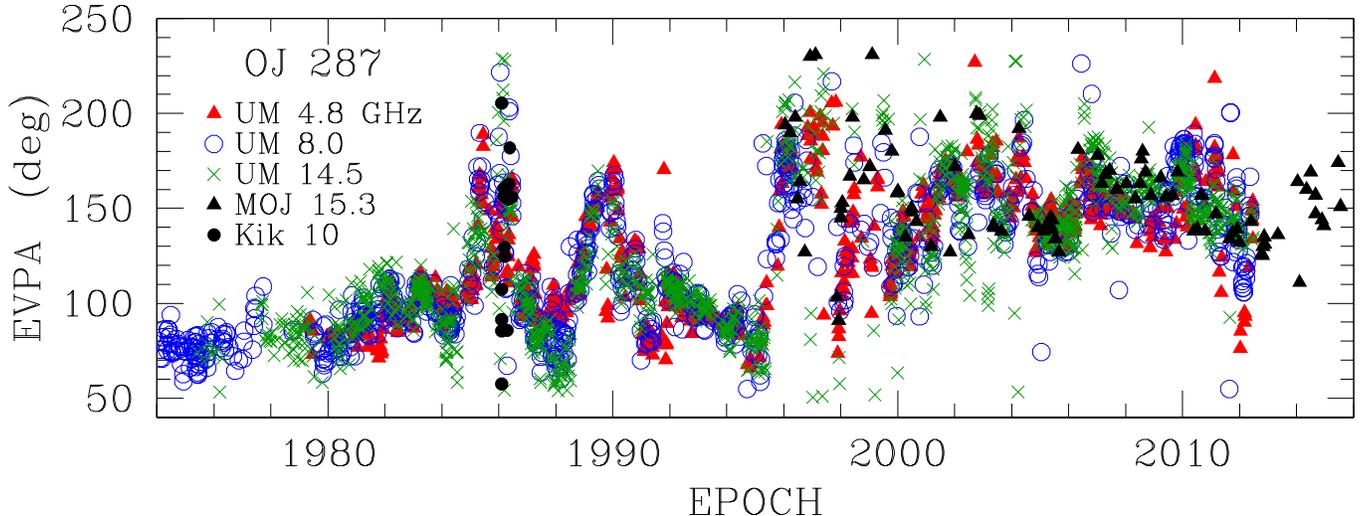}
\caption{
Archival EVPA data for OJ\,287 from UMRAO, MOJAVE, and \citet{KIM88}, 
presented in the range 50\deg--230\deg.
\label{evpaorig}}
\end{figure*}

%%%%%%%%%%%%%%%%%%%%%%%%%%%%%%%%%%%%%%%%%

We also use data from the MOJAVE
program\footnote{\url{http://www.physics.purdue.edu/MOJAVE}}
(Monitoring of Jets in Active Galactic Nuclei with VLBA Experiments),
which includes archival data back to 1995.  This is a continuing
program and for this paper we stop at 2016.0. MOJAVE uses the VLBA at
15.3 GHz.  An abbreviated version of the data analysis is as follows;
see \citet{LH05} for details. At each epoch we make images of Stokes
I, Q, and U, with pixel size 0.1 mas, and fit elliptical Gaussians
(circular if possible) to the I image, to find a set of `components'.
There typically is a bright component in Stokes I at the NE end of the
jet, and the center of this component is defined as the location of the
`core'. We cannot find similar components in the Q and U images because
polarization cancellation in close components can result in non-Gaussian
structures. Hence, we treat all Stokes parameters the same and find I,
Q, and U for the core by averaging over 9 pixels centered on the core.
The unit we use for I, Q, and U is Jy\,beam$^{-1}$.  The fractional
linear polarization is defined as $m=\sqrt{Q^2+U^2}/I$, and the EVPA is
calculated as $\mathrm{EVPA} = \xi/2 = (1/2)\tan^{-1} (U/Q)$.  In the
following we mix the flux densities (in Jy) from UMRAO with the specific
intensities (in Jy\,beam$^{-1}$) from the VLBA and use the symbol $F$
for all of them; the fractional linear polarization is called $P$, and
the product $PF$ is the linearly-polarized flux density. For the
VLBA data, $F$ and $PF$ are the flux densities of the compact core.

We have also used results obtained by other VLBA observers at 15.3 or
15.4~GHz, and placed in the VLBA archive. In these cases the data have
been reprocessed by the MOJAVE team, to make a homogeneous data set. The
combined points are typically a month apart, and by themselves would
be too infrequent for the rotation reversals we study in this paper,
but they are useful as a check on the UMRAO points. 

In addition to the UMRAO and VLBA data we use the results of
\citet{KIM88}, who made polarization observations of OJ\,287 at several
frequencies ranging from 9.0 to 10.5 GHz, for 6 months in 1986.  One of
our EVPA rotation reversals (Event A) occurred during their observing
period, and we include part of their data in our analysis. In
this period they observed on a daily basis, and this is important in
reducing ambiguity in Event A. The \citet{KIM88} data were taken
with the 45-m dish at Nobeyama. We use numerical values that were
found by digitizing the points in \citet{KIM88}, using the Dexter
tool.\footnote{\url{https://dexter.edpsciences.org/Dexterhelp.html}}

In Figure~\ref{evpaorig} we show the EVPA from the 5 archival data sets:
4.8, 8.0 and 14.5 GHz from UMRAO, 15.3/15.4 GHz from MOJAVE, and 9.0-10.5
GHz from \citet{KIM88}.  In the archives the data are listed in the range
$0\deg-180\deg$, but for Figure~\ref{evpaorig} we changed the range to
$50\deg-230\deg$, to better show the continuity of the points.  We
have ignored the Galactic Faraday Rotation towards OJ\,287 because
it is only of the order of 30 rad/$m^2$ \citep[e.g.,][]{rudnick83},
which rotates the 14.5 GHz EVPA by less than 1\deg.

In some regions of Figure~\ref{evpaorig} the EVPA varies smoothly,
but in others it is highly erratic. Therefore, we sought a smooth EVPA
curve by adding $\pm n\pi$ as appropriate.  \citet{K16} have derived some
procedures for this, based on a smoothness criterion, but we followed
the common practice of adding $\pm n\pi$ so that adjacent points differ
by less than 90\deg. However, we relaxed this rule when there was a
substantial time gap in the observations.  \citet{LBPP17} have made
a statistical study of how such gaps can affect the interpretation of
polarization data.  We also had a second criterion: make the curve fit
all frequencies as closely as possible.  This is important in reducing
ambiguity when one frequency has a data gap that can be filled by
another.

%However, there
%was one period about half a year long, 1985.8--1986.4, when the different
%frequencies could not be reconciled to one curve.  For that period we
%generated several solutions to the $n\pi$ problem, one by finding the
%single best curve for all the data, and the others by treating the
%frequencies independently. This is discussed in Section~\ref{s:eventA}.

%%%%%%%%%%%%%%%%%%%%%%%%%%%%%%%%%%%%%%%%%

%% FIGURE 2
\begin{figure*}
\centering
\includegraphics[width=0.9\textwidth,trim=0cm 0.6cm 1cm 0.5cm]{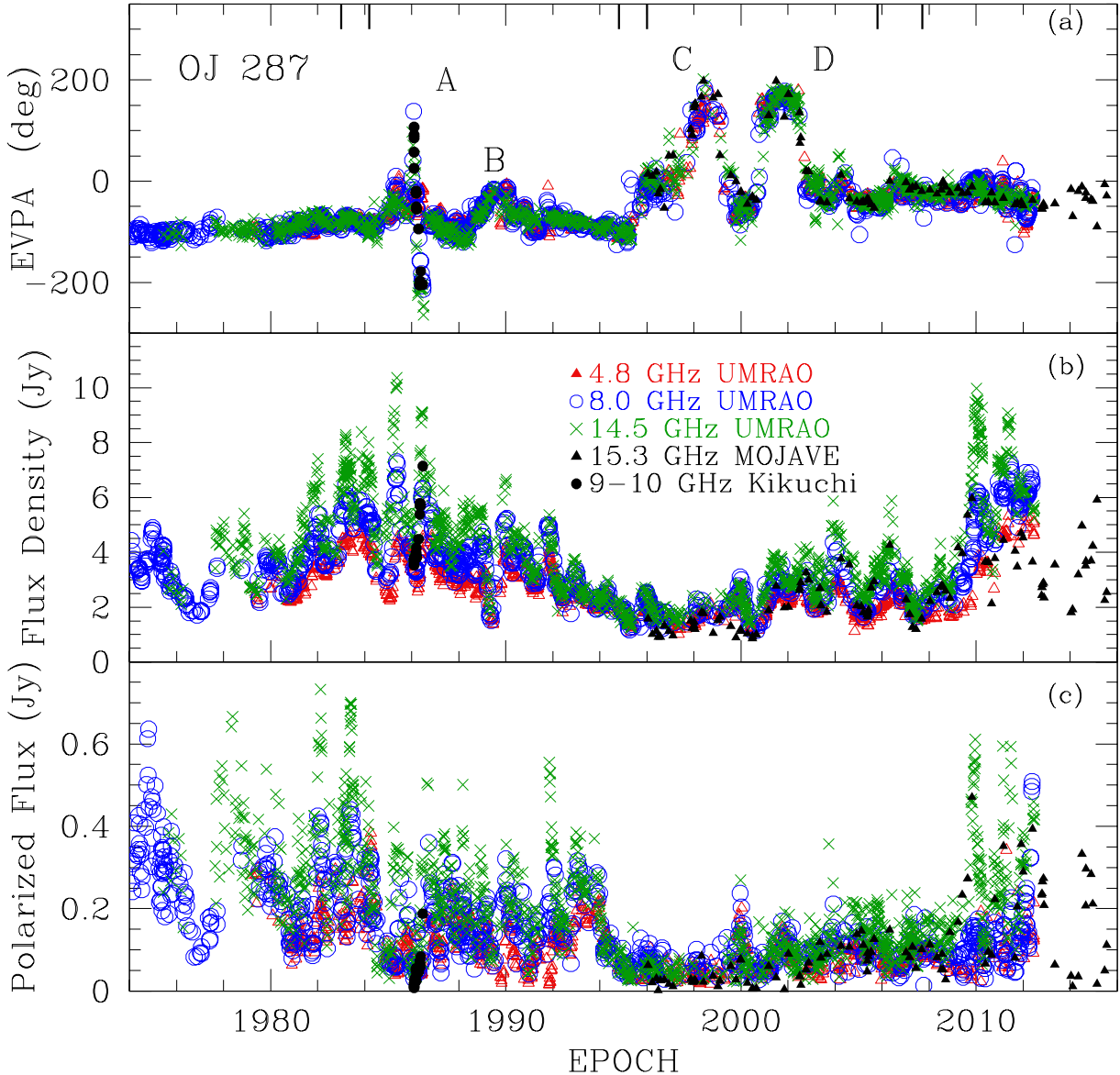} % fig TALL
\caption{
(a) The adjusted EVPA. Note the
different scales in Figures~\ref{evpaorig} and \ref{evpa7416}a. (b) Flux density from \citet{KIM88},
from the UMRAO archive, and from the MOJAVE archive for the core of
OJ\,287. (c) Polarized flux density. The bars on the top axis in
(a) indicate the epochs of the optical bursts that show a 12-year
quasi-periodicity. See text.
\label{evpa7416}
}
\end{figure*}

%%%%%%%%%%%%%%%%%%%%%%%%%%%%%%%%%%%%%%%%%

Figure~\ref{evpa7416}a shows the result we obtained for the smoothed EVPA
when we followed both criteria. In this Figure we identify three major
events and one minor event, labeled A, C, D, and B, respectively. Event
D is a smooth reversal; the EVPA swings CCW by about 200\deg, is stable
for roughly $1.5~y$ and then swings CW by about 160\deg. Event C is a
similar reversal with the same sign (CCW then CW) and similar amplitude,
but is narrower and appears to have a low-amplitude precursor. Event
A includes a sharp rotation reversal with the same sign as the others,
but with a larger CW swing. Event B has low amplitude and a different
shape. All these events are discussed in Section~\ref{s:reversals}.

%%%%%%%%%%%%%%%%%%%%%%%%%%%%%%%%%%%%%%%%%%%%%%

%% FIGURE 3
\begin {figure}
\centering
\includegraphics[scale=0.6]{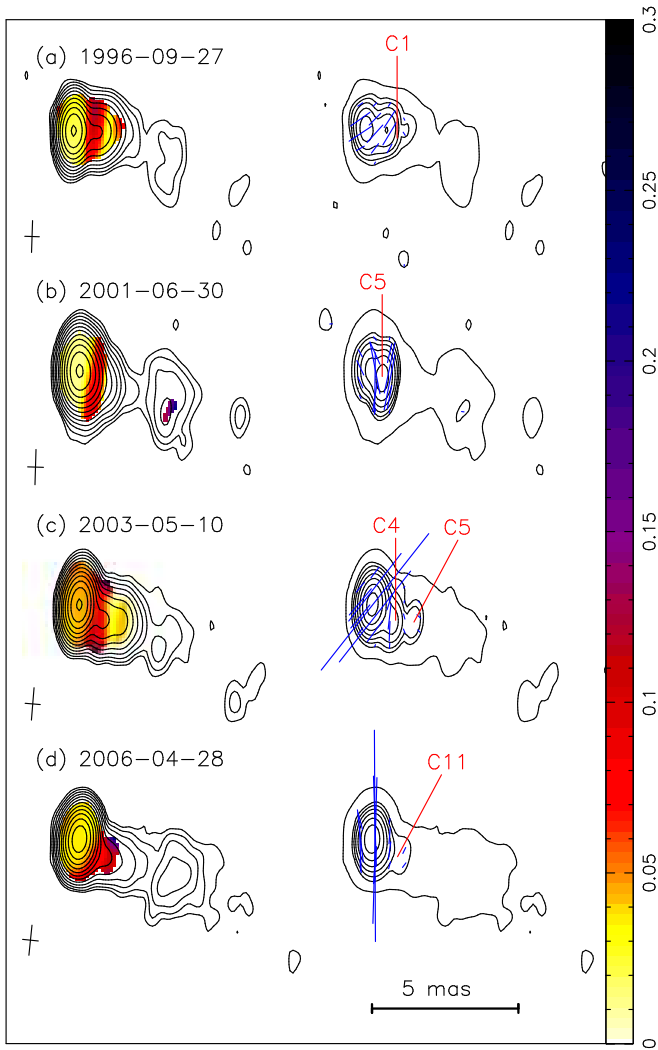}
\caption{
In each panel, Stokes I image is at left, with fractional linear polarization
in color.  The right-hand image is polarized flux $PF$ with an
additional contour that is the same as the lowest one for $I$. Sticks
show the direction of EVPA and the magnitude of $PF$: 50 mas per Jy.
Cross at lower left of each panel shows the restoring beam. The 
superluminal components C1, C5, C4, and C11 are indicated on the 
$PF$ images.
(a) Image from the VLBA archive, processed by MOJAVE,
(b) As in (a),
(c) As in (a),
(d) VLBA image from the MOJAVE program.  
\label{polimage}
}
\end{figure}

%%%%%%%%%%%%%%%%%%%%%%%%%%%%%%%%%%%%%%%%%%%%%%

We have two immediate results for OJ\,287 from Figure~\ref{evpa7416}. The
EVPA values from UMRAO and MOJAVE generally lie close together, and
so the EVPA data obtained with a 26-m dish \textit{are usually a good
proxy for VLBA measurements for the core alone}.  This assertion can be
tested by examining the MOJAVE polarization images \citep{Lis18}. In
most of them (51/59) the core is clearly the strongest component in
$PF$, and so the polarization of the total source is similar to that of
the core.  In 8/59 images a secondary component is stronger. However,
they are not distributed uniformly in time, but all occur during Events
C and D.  Figure~\ref{polimage} shows four examples of the images. In
each panel the left-hand image shows the contours of Stokes I, and the
linear polarization fraction is in color. The right-hand image shows the
contours of $PF$, the linearly polarized flux density, with an additional
contour that is the same as the lowest contour of the Stokes I image.
In Figures~\ref{polimage}c and \ref{polimage}d the cores are stronger
than the secondary components in $PF$, but in Figure~\ref{polimage}b the
core is weaker and in Figure~\ref{polimage}a the core and the secondary
component have similar strength. We discuss this further in Sections 4.2
and 4.3.

%, where we show that the EVPA of the core is similar to that of
%the tkotal source even when the polarized flux density is dominated by a
%secondary component. Since the inner jet is defined by the location of the
%secondary component, we see that the core has the same EVPA as the PA of
%the inner jet \citep{AMJ12, C17}, consistent with the assertion above.

The second result from Figure~\ref{evpa7416}a is that we can assume
that the EVPA is largely frequency-independent over the range 4.8
-- 15.4 GHz. This is consistent with most of the data.  However, the
frequency-independence is violated in Event A, from 1985.9 to 1986.5, when
the points at 4.8 GHz are separated from those at the other frequencies,
as discussed in Section~\ref{s:eventA}.

The $n\pi$ adjustments that convert Figure~\ref{evpaorig} into
Figure~\ref{evpa7416}a were made by hand. We also used an algorithm
similar to that of \citet{K16} that tests every point against the
previous one, and adds $\pm n\pi$ as necessary to keep the difference
below 90\deg.  This is an automatic procedure that does not allow for
any special considerations at a data gap. We did this for points at
the different frequencies being treated separately, and also for all
the points being used together.  For the latter case, the results were
similar to the non-automatic solution shown in Figure~\ref{evpa7416}a.

Figure~\ref{evpa7416}b shows the flux density $F$ of OJ\,287 at the
five frequencies. The MOJAVE values are for the core, but the others
are total-flux measurements made with a single large dish.  At most
epochs OJ\,287 has an ``inverted" spectrum, with $S_{14.5} > S_{4.8}$,
like many AGN \citep{KNK99, FAZ16}. The 15.3 GHz flux densities for
the core are usually well below the 14.5 GHz values for the total
source, and show that the jet makes a substantial contribution to
the total flux density.  This is especially noticeable after 2010.
Figure~\ref{evpa7416}c shows the polarized flux density, $PF$, which
will be important in the discussion of models for the EVPA rotations.

Tables~\ref{UMRAOtable}, \ref{MOJAVEtable} and \ref{KIKUCHItable}
contain all the points in the adjusted data sets, as shown in
Figure~\ref{evpa7416}. These tables contain 856 points at 4.8 GHz,
917 at 8.0 GHz, 1,207 at 14.5 GHz, and 93 at 15.3 GHz.
The archival data
can be reconstructed from Tables~\ref{UMRAOtable} and \ref{MOJAVEtable}
by constraining each EVPA point to lie in the range 0\deg~to 180\deg,
by adding $n\pi$ as needed. Table~\ref{KIKUCHItable} contains the
19 Kikuchi points at 9.0-10.5 GHz, found by digitizing the plots in
\citet{KIM88}. In this process the epochs differ slightly among the
points for the EVPA, $F$, and $P$; and the mean epoch is shown in column
1 of Table~\ref{KIKUCHItable}.

%%%%%%%%%%%%%%%%%%%%%%%%%%%%%%%%%%%%%%%%%%%%%%%%%%%%%%%%%%%%%%%

\newcommand{\n}{\nodata} 
\begin{deluxetable}{crrrr}
\tablecolumns{5} 
\tabletypesize{\scriptsize} 
\tablewidth{0pt}  
\tablecaption{\label{UMRAOtable}UMRAO Single Dish Data}  
\tablehead{\colhead{Epoch} & 
   \colhead{$\nu$} & \colhead{$F_\mathrm{tot}$}  & 
\colhead{$m$} &\colhead{EVPA}   \\  
\colhead {(y)} &\colhead{(GHz)} &\colhead{(Jy)} & \colhead{(\%)} &\colhead{(\arcdeg)}  \\  
\colhead{(1)} & \colhead{(2)} & \colhead{(3)} & \colhead{(4)} & \colhead{(5)}} 
\startdata 
 1971.336 & 8.0 & 3.71 & 5.5 & $-$90.3 \\ 
 1972.127 & 8.0 & 5.93 & 2.0 & $-$46.4 \\ 
 1972.143 & 8.0 & 4.77 & 5.0 & $-$86.5 \\ 
 1972.217 & 8.0 & 6.27 & 2.4 & $-$48.1
\enddata 
\tablecomments{Columns are as follows:  (1) observation epoch, (2) observation frequency in GHz, (3) total flux density in Jy,  (4) fractional linear polarization in per cent, (5) adjusted electric vector position angle in degrees. This table is published in its entirety in the electronic edition of the Journal; a portion is shown here for guidance regarding form and content.}
\end{deluxetable} 

%%%%%%%%%%%%%%%%%%%%%%%%%%%%%%%%%%%%%%%%%%%%%%%%%%%%%%%%%%%

\begin{deluxetable}{cccr}
\tablecolumns{4}
\tabletypesize{\scriptsize}
\tablewidth{0pt}
\tablecaption{\label{MOJAVEtable}MOJAVE 15 GHz VLBA Core Feature Data}
\tablehead{\colhead{Epoch} & \colhead{I} &\colhead{$m$} &  \colhead{EVPA}  \\
\colhead {(y)}  &\colhead{(Jy beam$^{-1}$)}   &\colhead{(\%)}   &\colhead{(\arcdeg)}  \\
\colhead{(1)} & \colhead{(2)} & \colhead{(3)} & \colhead{(4)}}
\startdata
1996.049 & 1.57 & 4.0 & 14    \\
1996.222 & 1.01 & 2.1 & 10    \\
1996.402 & 1.13 & 2.7 & 18    \\
1996.474 & 0.92 & 0.2 & $-$25
\enddata
\tablecomments{Columns are as follows: (1) observation epoch, (2) Stokes I intensity in Jy/beam, (3) fractional linear polarization in per cent, (4) adjusted electric vector position angle in degrees. This table is published in its entirety in the electronic edition of the Jornal; a portion is shown here for guidance regarding form and content.}
\end{deluxetable}

%%%%%%%%%%%%%%%%%%%%%%%%%%%%%%%%%%%%%%%%%%%%%%%%%%%%%%%%%%%%%%%

\begin{deluxetable}{lrrr} 
\tablecolumns{4} 
\tabletypesize{\scriptsize} 
\tablewidth{0pt}  
\tablecaption{\label{KIKUCHItable}Kikuchi 10 GHz Radio Data}  
\tablehead{\colhead{Epoch} &   \colhead{$F_\mathrm{tot}$}  &\colhead{$m$}  & \colhead{EVPA} \\  
\colhead {(y)} &\colhead{(Jy)}  &\colhead{(\%)}  &\colhead{(\arcdeg)} \\  
\colhead{(1)} & \colhead{(2)} & \colhead{(3)} & \colhead{(4)}} 
\startdata 
1986.092        &       \n      &       \n      &       107.3   \\
1986.094        &       \n      &       \n      &       85.4    \\
1986.097        &       0.33    &       3.53    &       91.4    \\
1986.101        &       0.23    &       3.67    &       57.4    \\
1986.105        &       0.47    &       3.76    &       25.4    \\
1986.111        &       0.75    &       3.80    &       \n      \\
1986.181        &       \n      &       \n      &       $-$19.5 \\
1986.185        &       1.33    &       3.80    &       $-$23.5 \\
1986.187        &       1.36    &       3.98    &       $-$23.5 \\
1986.193        &       0.59    &       4.17    &       $-$50.5 \\
1986.198        &       1.05    &       3.80    &       $-$55.5 \\
1986.201        &       1.44    &       3.85    &       \n      \\
1986.288        &       1.02    &       4.48    &       $-$94.3 \\
1986.335        &       \n      &       \n      &       $-$205.3        \\
1986.338        &       1.35    &       5.38    &       $-$204.3        \\
1986.342        &       1.03    &       5.70    &       $-$197.3        \\
1986.347        &       1.14    &       5.79    &       \n              \\
1986.379        &       1.46    &       5.70    &       $-$178.2        \\
1986.467        &       2.63    &       7.14    &       $-$204.1        
\enddata
\tablecomments{Columns are as follows: (1) observation epoch,  (2) 
total flux density in Jy, (3) fractional linear polarization
in per cent, (4) adjusted electric vector position angle in degrees.}
\end{deluxetable}

%%%%%%%%%%%%%%%%%%%%%%%%%%%%%%%%%%%%%%%%%%%%%%%%%%%%%%%%%%%%%%%

\section{Are the Rotations with a Reversal Spurious?}
\label{s:spurious}

%%%%%%%%%%%%%%%%%%%%%%%%%%%%%%%%%%%%%%%%%%%%%

%% FIGURE 4
\begin{figure*}
\centering
\includegraphics[width=0.9\textwidth,trim=0cm 0.5cm 1cm 0.5cm]{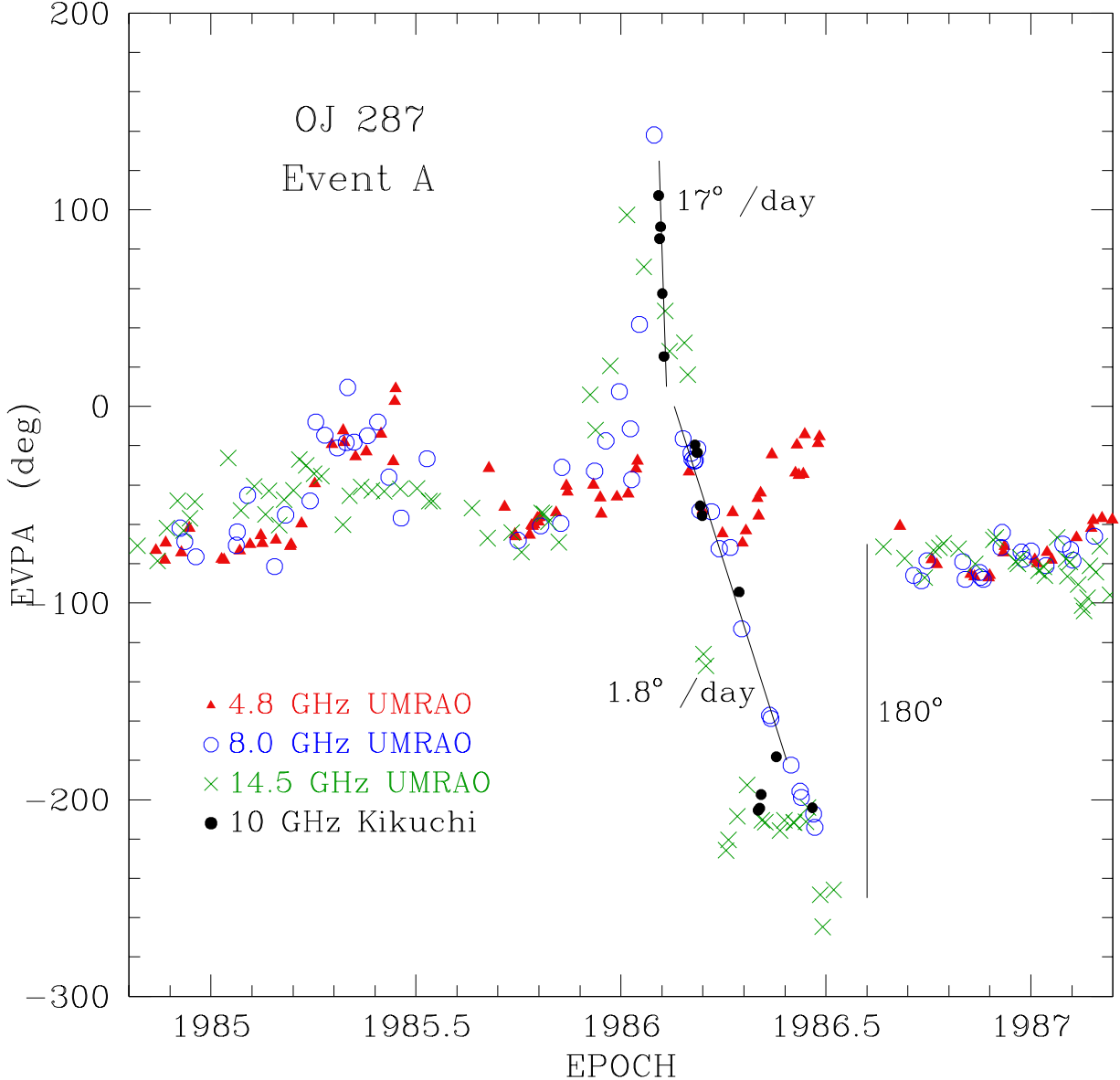}
\caption{
EVPA for Event A. The step at 1986.52 is cosmetic, and the plot
shows that there is a stable EVPA near -70\deg~that exists before and
after the event. See text.
\label{fig:eventA}
}
\end{figure*}

%%%%%%%%%%%%%%%%%%%%%%%%%%%%%%%%%%%%%%%%%%%%%%

\citet{LJ16} have emphasized that measured EVPA rotations can be spurious
for two reasons: they can be generated by a random walk process, and they
can be both generated and destroyed by statistical noise.  This has also
been discussed by e.g., \citet{JRA85, M14, K16, KBP17}.  In this Section
we ask if these effects can be at work in our observations.  We believe
that they are not, because the probability of a random large double
rotation with a reversal must be much smaller than the probability of a
single rotation, but in OJ\,287 we see three large reversals in 42 years,
with no similar single rotations. In addition, the four reversals that
we see are all in the same direction: CCW then CW. This alone reduces
the probability that the rotations are random by an order of magnitude.

\citet{JRA85} first estimated the probability that a large EVPA rotation
could be due to a random process. They considered a source that consisted
of  turbulent cells with random polarizations, and evolved the system
by changing one cell per time step. With Monte-Carlo calculations, they
found a rather high probability of a large rotation; with appropriate
assumptions the probability of a rotation of 180\deg~ or greater was
as much as 0.3. For our purposes we need to multiply this estimate by
the probability that the next rotation has a similar amplitude and the
opposite sign, occurs shortly after the first one, and is isolated; i.e.,
there is no third rotation for a substantial period. This appears to call
for a Monte-Carlo calculation, which is beyond the scope of this paper.
However, it is clear that each of these factors will appreciably
reduce the overall probability for the observed double rotations to
arise by chance, compared to the probability for a single large rotation.

Another factor affecting the probability is that the
rotations occur at the same time with independent observations at three
frequencies. We have coincident events, and this greatly reduces the
probability that they are due to random noise. But it may not
reduce the probability that they are due to random walks, since the 
turbulent cells may be frequency-independent. For this to be the case,
however, opacity effects must be negligible. 

%%%%%%%%%%%%%%%%%%%%%%%%%%%%%%%%%%%%%%%%%%%%%%%%%%

\section{The Rotation Reversals}
\label{s:reversals}
\subsection{Event A}
\label{s:eventA}                    % EVENT A

%%%%%%%%%%%%%%%%%%%%%%%%%%%%%%%%%%%%%%%%%%%%%%%%%%%%%%

%% FIGURE 5
\begin{figure*}
\centering
\includegraphics[width=0.9\textwidth,trim=0cm 0.5cm 0cm 0.5cm]{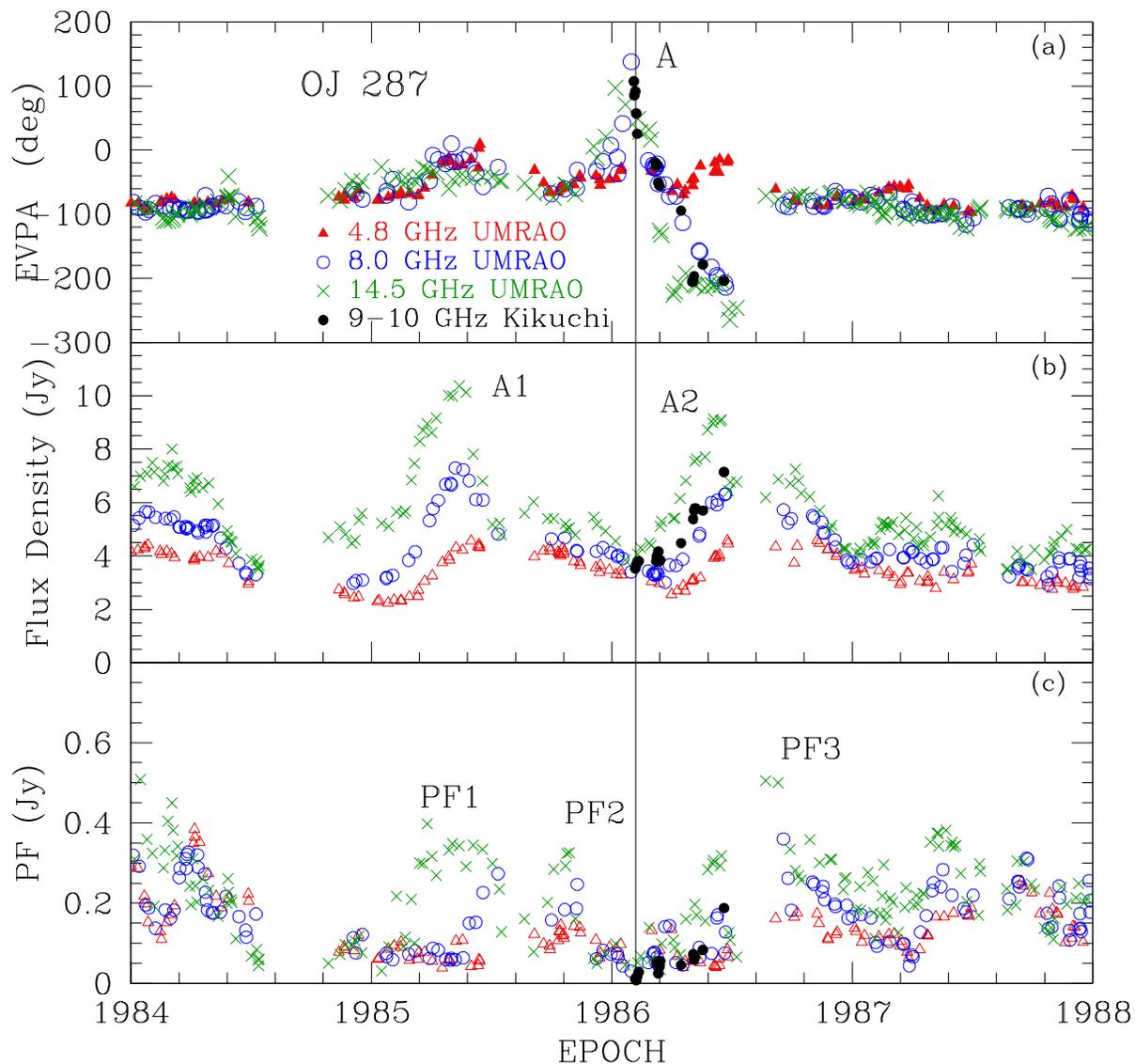}
\caption{
Expanded view of Figure~\ref{evpa7416}, 1984-1988.
The line at 1986.1 connects the
14.5 GHz reversal with the minimum in $PF$.
\label{evpa8488}}
\end{figure*}

%%%%%%%%%%%%%%%%%%%%%%%%%%%%%%%%%%%%

In this Section we describe the principal rotation events that
are seen in Figure~\ref{evpa7416}. We first discuss Event A, because
it is bracketed by outbursts in total and polarized flux density, and
this motivates the model we describe later. We then discuss, in order
of complexity, events D, C, and B.

Our result for the EVPA of Event A is shown in
Figure~\ref{fig:eventA} and was obtained by following our two connection
criteria: generally keep adjacent points less than 90\deg~ apart except
where there is a large time gap, and keep all frequencies on the same
curve to the extent possible. To make the curve we first noted that the
first five Kikuchi points, marked 17\deg/day, form a steep line that is
unambiguous, as are the 14 8.0 GHz points that are indicated with the
line marked 1.8\deg/day. The two lines fit together well and define the
main structure of the EVPA curve. The other points for 8.0 and 10 GHz
then connect as shown. The points for 4.8 GHz show no evidence for the
steep CW rotation seen at the other frequencies, and we dropped the
requirement that the 4.8 GHz points had to fit in with the others.
The 14.5 GHz points from 1986.2 to 1986.5 do not fit well with the
others, and we placed them close to the 10 GHz line, since 10 GHz is
the nearest frequency. This is arbitrary, and raising them by 180\deg~
would place them close to the 4.8 GHz points.  As we discuss later,
in Section~\ref{s:stokes_vectors} in connection with the two-component
model, these differences might result from the different behaviors of
the polarized flux, at the different frequencies.

%%%%%%%%%%%%%%%%%%%%%%%%%%%%%%%%%%%%%%%%%%%

%% FIGURE 6
\begin{figure*}
\centering
\includegraphics[width=0.9\textwidth,trim=0cm 0.5cm 0cm 0.5cm]{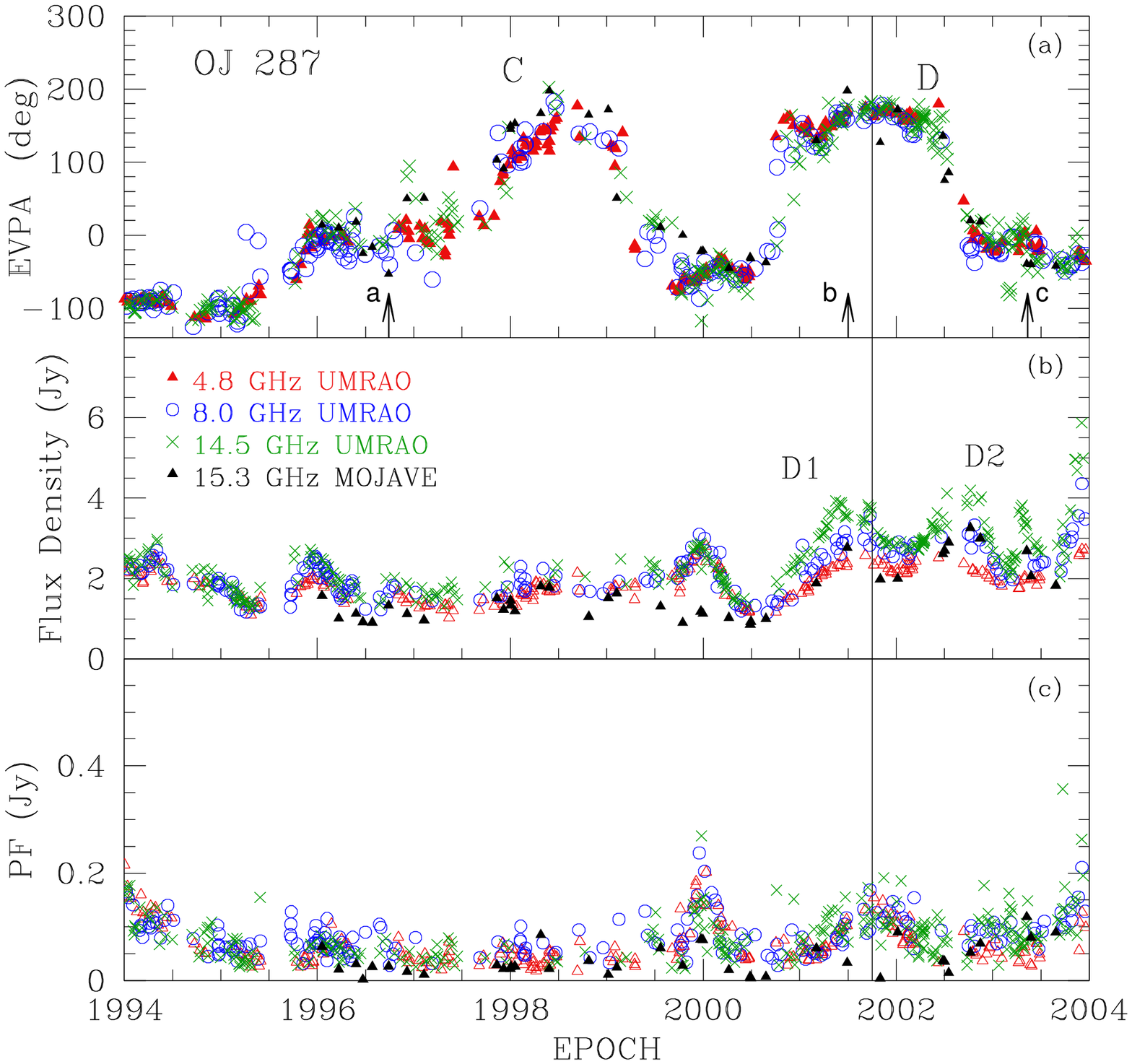}
\caption{
Expanded view of Figure~\ref{evpa7416}, 1994-2004. The line at 2001.75
connects the EVPA reversal in (a) and the peak $PF$ in (c).
Arrows on the abscissa of (a) indicate the epochs of the 
corresponding images in Figure~\ref{polimage}.
\label{evpa9404}
}
\end{figure*}

%%%%%%%%%%%%%%%%%%%%%%%%%%%%%%%%%%%%%%%%%%%

Figure~\ref{fig:eventA} shows that Event A had a CCW EVPA rotation
of about 180\deg, followed by a CW rotation of roughly 360\deg. The EVPA
before Event A was about --60\deg, and roughly --260\deg~after it. But
--260\deg~is the same as --80\deg, and so we inserted a step of +180\deg~
at 1986.52 for cosmetic purposes, to make it easy to see that the EVPA
was approximately the same before and after the event.

Figure~\ref{evpa8488} gives an extended view of Event A.
with the three panels showing the EVPA, 
the flux density, and the linearly--polarized flux density.  Two large
outbursts in flux density, A1 and A2, bracket the EVPA event. They show
the normal evolution of emission from an expanding synchrotron cloud,
with lower frequencies delayed and reduced in intensity.  A weak double
rotation in EVPA, with a reversal, occurs at the same time as the peak
of outburst A1.  The strong EVPA Event A occurs during the tail of A1
and the rise of A2, and the reversal itself occurs at the time of the
$F$ and $PF$ minimum between A1 and A2.

The polarized flux $PF$ has a complex pattern in this interval.  We only
discuss the highest frequency, 14.5 GHz.  $PF$ has a strong peak, PF1,
at 1985.4, at the time of the weak EVPA reversal, and it has a deep
minimum at 1985.6, when the EVPA has almost returned to its baseline
value. The $PF$ has peaks PF2 and PF3 bracketing Event A, and is in a
deep minimum through much of Event A.  The CCW swing in EVPA from 1985.8
to 1986.1 occurs during the tail of A1, and the CW swing from 1986.1 to
1986.4 occurs during the rise of A2. Thus A1 itself, or at least
its tail, is polarized with CCW rotation, and similarly the rise of A2
has CW rotation. The observed reversal in rotation occurs when A2 begins to
dominate the total $PF$, at 1986.1. The deep minimum in $PF$ at that time
implies that the two components have EVPAs that are roughly 90\deg~apart.

The EVPA swings in Event A are of order 180\deg~or more, and they cannot
be due to two variable sources with fixed EVPA, nor to the evolution
of optical depth, since both of these give a maximum swing of 90\deg.
In Section~\ref{s:stokes_vectors} we present a model consisting of a
steady polarized component combined with a variable component that has a
rotating EVPA. If the components have similar amplitudes then when the
EVPAs are nearly perpendicular the net $PF$ will have a minimum, and the
EVPA can have a rapid swing.  This model can explain many of the observed
features of the EVPA reversals in OJ\,287.

\subsection{Event D}           %% NEWD
\label{s:eventD}

Event D is shown in Figure~\ref{evpa9404}. It is bracketed by modest
outbursts D1 and D2, similar to the way in which Event A is bracketed by
A1 and A2.  The EVPA has a rapid CCW swing in late 2000 at a rate $\sim
1.7\deg~ d^{-1}$; the rate is not uniform.  After the CCW swing the EVPA
is nearly steady for about 1.5 years and then has another rapid swing,
this time CW, at  the rate $\sim -0.8\deg d^{-1}$.  These rotations
in EVPA occur during the rise of D1 and  D2. As with Event A they are
greater than 90\deg~and cannot be solely due to evolution of  optical
depth, or to a combination of two variable sources with fixed EVPA.

The $PF$ for Event D has a peak near 2001.75, in the middle of the
steady period for the EVPA, and the $PF$ has  minima during the rapid
EVPA swings. This is different from the behavior in Event A, where
the EVPA reversal at 1986.1 occurs during a $PF$ minimum. (See
Figure~\ref{evpa8488}). This will be discussed in terms of the
two-component model in Section~\ref{s:optical}.

The arrows on the abscissa of Figure~\ref{evpa9404}a correspond to the
epochs for the images in Figure~\ref{polimage}.  The $PF$ images show the
core and one or two secondary jet components to the W or SW.  These jet
components are Nos. 1, 5, 4 and 11 in the MOJAVE list \citep{Lis13, Lis16}
and are labeled C1, C5, C4 and C11 in Figure~\ref{polimage}.\footnote{The
component labeled C5 is probably a blend of C5 with C10, the slow
component near the core; and  C11 is probably a blend of C11 with C9.}

These four components are all superluminal, and C4 is the fastest one,
with $\beta_\mathrm{app}=15$.  We are especially interested in C5, because
it is intimately connected to Event D.  C5 is moving at the rate $\mu=0.54
\pm 0.07$~mas\,y$^{-1}$ in the direction $\mathrm{PA}=-103\deg$. It is
not moving radially but projects back close to the core, and was near
the location of the core around 1999--2000, assuming that it was in
uniform motion \citep{Lis16}.

%%%%%%%%%%%%%%%%%%%%%%%%%%%%%%%%%%%%%%%%%%%%%%%%%%%%

%%FIGURE 7
\begin{figure*}
\centering
\includegraphics[width=0.9\textwidth,trim=0cm 0.5cm 0cm 0.5cm]{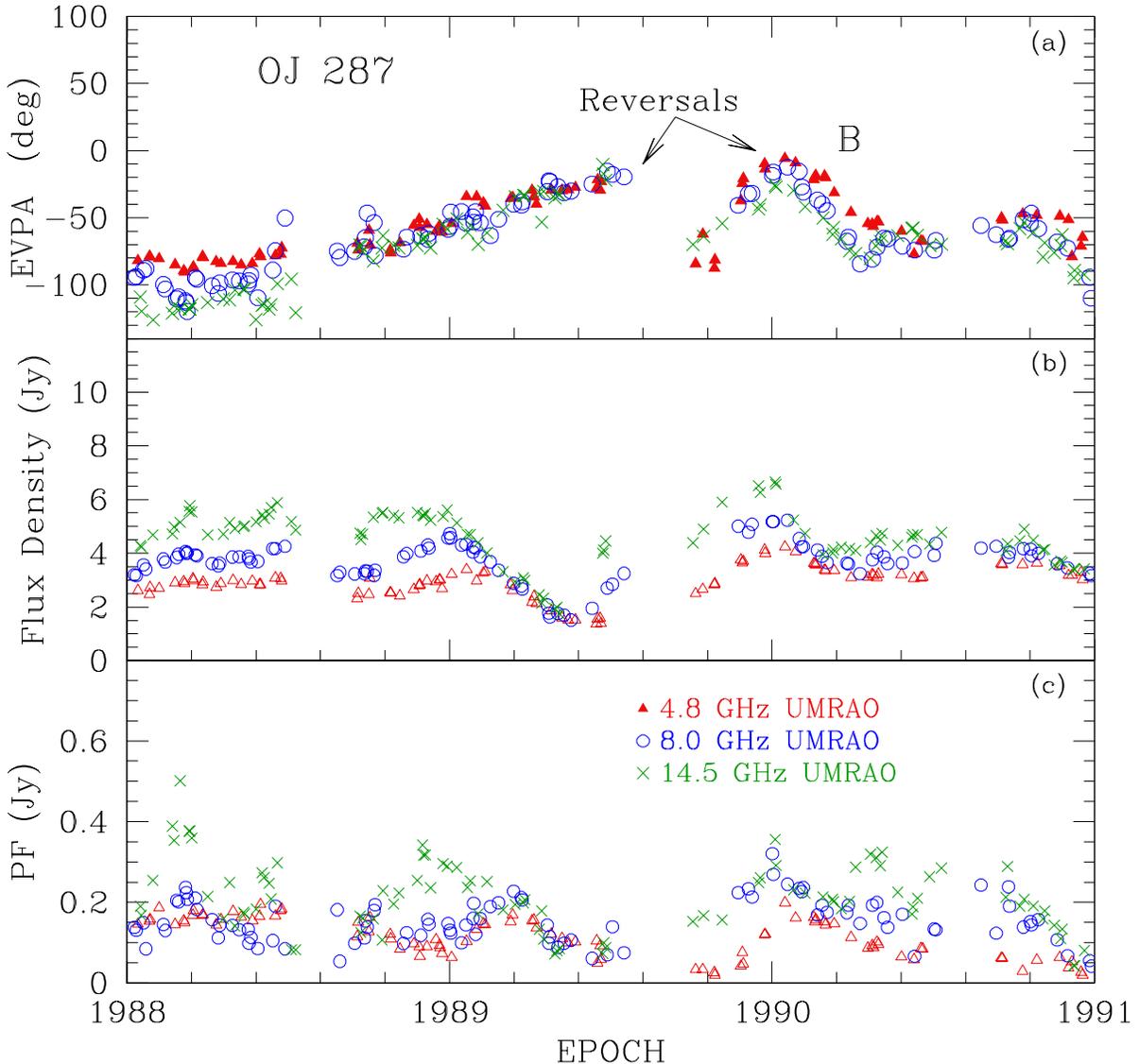}
\caption{Expanded view of Figure~\ref{evpa7416}, 1988-1991.
\label{evpa8891}
}
\end{figure*}

%%%%%%%%%%%%%%%%%%%%%%%%%%%%%%%%%%%%%%%%%%%%%%%%%%%%

In Figure~\ref{evpa9404} both $F$ and $PF$, at 14.5 GHz (green crosses in
panels b and c), begin to increase around 2000.5. $F$ rises to nearly 4~Jy
by 2001.4, while $PF$ continues to rise until nearly 2002.  The $PF$ image
in Figure~\ref{polimage}b shows that the $PF$ rise is due to component C5,
which dominates the image, and
presumably first became visible around 2000.5 when the total flux
density started to increase.  In this case the simultaneous increase in
$F$ is also due to C5, although this is not so obvious in the total flux
image in Figure~\ref{polimage}b.

%The 15.3
%GHz MOJAVE flux in Figure~\ref{evpa9404}b rises less than the 14.5 GHz
%UMRAO flux, because the MOJAVE values are for the core, while most of the
%increase is for the downstream component C5. In D1 the three UMRAO curves
%are displaced vertically as is common for a flux outburst, and is usually
%explained in terms of the evolution of an expanding synchrotron source.
%However, we suggest instead that the evolution of F in D1 is mainly
%controlled by the emergence of C5 from a $\tau=1$ region, and not by
%the expansion of a synchrotron cloud.

%Component C11, seen in Figure~\ref{polimage}d, had a different behavior;
%its flux density increased well before its separation from the core
%became visible \citep{C17}.  Thus, its appearance was controlled by the
%angular resolution of the VLBA, and not by opacity effects.

The rise of the outburst D1 in Figure~\ref{evpa9404}b has CCW EVPA
rotation, but the subsequent decline has a steady EVPA, as seen in
Figure~\ref{evpa9404}a.  D2 has CW rotation and the combination of the
core and the two bursts starts to rotate CW when D2 starts to dominate
the flux density.  This happens around 2002.3. $PF$ has a minimum then
because, according to the model in Section~\ref{s:double_rotation},
the sum of the Stokes vectors for the three components becomes
small. At that time the phase of the sum, $\xi$, can sweep rapidly,
and so the $\mathrm{EVPA} = \xi/2$ also sweeps rapidly. 

%%%%%%%%%%%%%%%%%%%%%%%%%%%%%%%%%%%%%%%%%%

%%FIGURE 8
\begin{figure*}
\centering
\includegraphics[width=0.95\textwidth,trim=0cm 9cm 0.5cm 0.5cm]{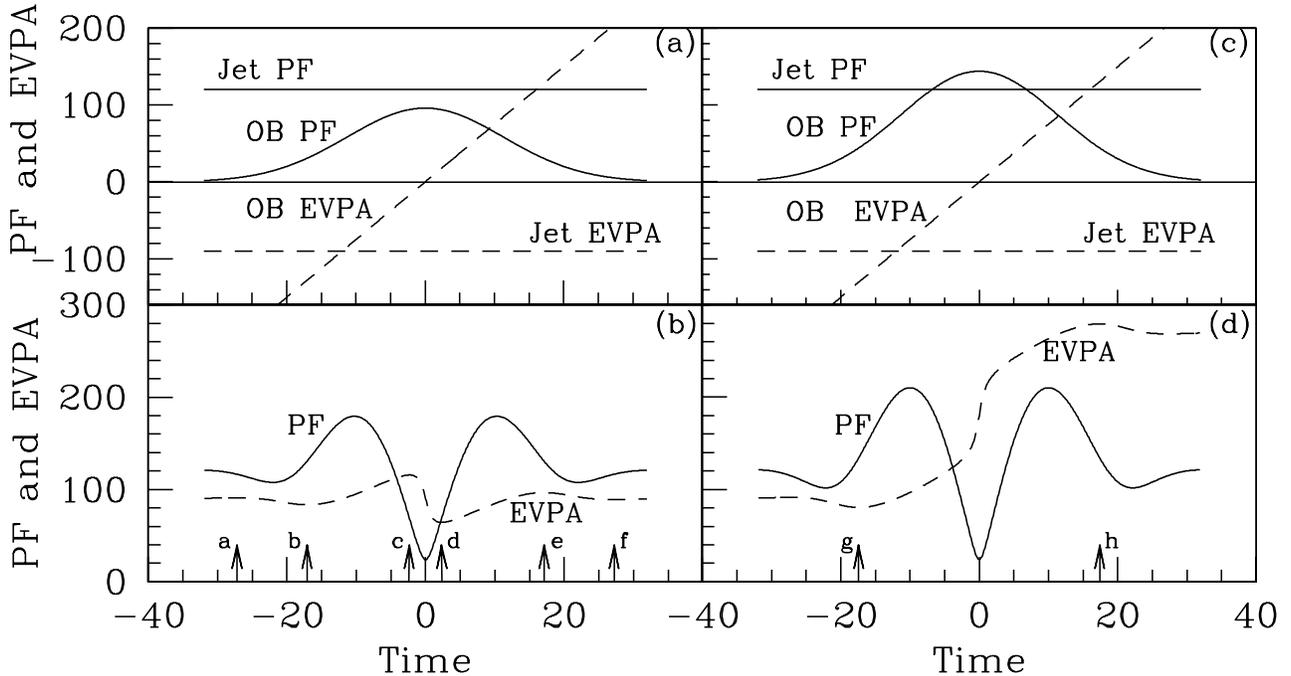} 
\caption{Combination
of an outburst (OB) whose polarized flux has a Gaussian time
dependence and an EVPA that rotates CCW, with a steady jet component.
The polarized fluxes are shown with solid lines, and the EVPAs are
dashed.  (a) The jet is stronger than the peak of the Gaussian.
(b) the resultant of the two components in (a), obtained by summing
their Stokes parameters. (c) As in (a) but the Gaussian is stronger
than the jet.  (d) As in (b).  Arrows on the abscissae of panels (b)
and (d) correspond to the labeled dots in Figure~\ref{fig:stokes}.
\label{fig:model_4}
}
\end{figure*}

%%%%%%%%%%%%%%%%%%%%%%%%%%%%%%%%%%%%%%%%%%

\subsection{Event C}           %% NEWC
\label{s:eventC}

The events in Figure~\ref{evpa9404} start at 1995.5 with a CCW swing in
EVPA of about 100\deg, coincident with small bursts in $F$ and $PF$.
They are presumably due to the emergence of C1 from the core; C1
is seen 1.3 years later in Figure~\ref{polimage}a. The epoch of this
image is shown in Figure~\ref{evpa9404}a with the arrow marked ``a".
The left-hand image in Figure~\ref{polimage}a shows that C1 is highly
polarized, but the right-hand image shows that the core, while weakly
polarized, has more polarized flux density.  The flux burst at 1996.0 does
not show the common high-to-low frequency evolution, and we can ignore the
possibility of EVPA changes due to optical depth effects. The 100\deg~EVPA
swing could be due to a combination of variable sources that have  fixed
EVPA, but it could also be due to sources with a variable EVPA.  In the
scenario presented in Section~\ref{s:stokes_vectors} the new component
C1, responsible for the flux burst at 1996.0, would have an EVPA with
CCW rotation.

In 1997.2--1998.5 Event C has a CCW EVPA swing of about 160\deg, followed
by a CW swing of about 200\deg. $F$ and $FP$ change little during the event,
The large EVPA swing is about the same in the UMRAO and MOJAVE points,
however, suggesting that the EVPA rotation is in the core. Note that the
$PF$ values from MOJAVE are very low, implying that the errors in EVPA are
high, and so the individual MOJAVE EVPAs should be treated with caution.

The MOJAVE EVPA points in Event C, starting near 1996.0, were plotted
in a different way by \citet{C17}, who did not show the reversal at
1998.5. This resulted from large gaps in the MOJAVE data; and without the
closely-spaced points from UMRAO it is difficult to obtain the correct
curve. The earlier work by \citet{Hom02} also shows a different curve,
and may have similarly suffered from the lack of closely-spaced points.

%%%%%%%%%%%%%%%%%%%%%%%%%%%%%%%%%%%

\subsection{Event B}

Figure~\ref{evpa8891} shows an expanded view of the period 1988-1991.
Again there is a sun gap, at 1989.6, that interferes with the
interpretations. Event B includes two EVPA reversals, one near 1989.5
that is preceded by a shallow CCW rotation of about 100\deg~and
followed by what is probably a steep CW rotation of at least 60\deg.
Unfortunately, the data are missing for this last rotation. The
second reversal, near 1990.0, is symmetric. These two reversals have
the sign that is common to all the reversals in OJ\,287, CCW then CW.
Detailed modeling is needed to investigate event B.

%%%%%%%%%%%%%%%%%%%%%%%%%%%%%%%    RSV

\section{Two-Component Model and Rotating Stokes Vectors} 
\label{s:stokes_vectors}

In this Section we present a two-component model that can reproduce many
of the polarization features seen in the preceding Sections.  
Two-component models have frequently been used to describe
polarization events.  \citet{B82} analyzed polarization changes due
to relativistic aberration, and compared them to changes that can be
produced by a non-relativistic two-component model.  \citet{HB84} used
a multi-parameter two-component model for OJ\,287, with both components
having  variable spectrum and polarization, needed to match the observed
time-dependent spectrum, flux density, polarization fraction, and
EVPA. More recently, \citet{BDA17} used a two-component model to explain
flaring activity in PKS 1510-089. \citet{Vil10} developed a two-component
model that is similar to ours; it is discussed in Section~\ref{s:optical}.

Our model
has a steady component that we call the jet, and a time-dependent
component that we refer to as the outburst. The outburst has a
Gaussian shape with truncated tails, and an EVPA that rotates uniformly.
The amplitude ratio of the two components, and the EVPA rotation rate and
phase, are picked so that the results mimic some of the observations. A
single rotation is modeled with one outburst. A double rotation, with a
reversal, can be modeled with two successive outbursts that have opposite
senses of EVPA rotation. The direct observation of outbursts A1 and A2,
seen in Figure~\ref{evpa8488}, motivates this model.

Figure~\ref{fig:model_4}a shows our model for the case where the peak of
the Gaussian outburst is weaker than the jet $\rm (Gaussian/jet=0.8)$;
and the other case, with the Gaussian stronger than the jet $\rm
(Gaussian/jet=1.2)$, is shown in Figure~\ref{fig:model_4}c.  The relative
size of the two components is important, for it controls the details
of the EVPA of the combination.  The Gaussians are truncated at $t=-32$ and
$t=32$. The rotation rate for the EVPA of the outbursts is
+7.5\deg~per unit time step. The EVPA of the jet is 90\deg~ from that
of the outburst at its maximum, at $t=0$.

Figures~\ref{fig:model_4}b and \ref{fig:model_4}d show the results
of combining the two components. In both cases $PF$ has a deep
minimum where the Gaussian and the jet have similar amplitudes, and
where their EVPAs are nearly perpendicular. But the resultant EVPAs
behave differently.  In Figure~\ref{fig:model_4}b the EVPA curve of the
combination has 6 extrema, or reversal points, at epochs indicated by
the arrows on the abscissa. There is a weak maximum at point a at 
$t=-27.2$, then a shallow CW swing to point b at $t=-17.1$, where
the rotation direction reverses, a CCW swing to point c at $t=-2.3$,
then another reversal and a rapid CW swing to point d at $t=+2.3$,
where the process repeats in reverse.  In Figure~\ref{fig:model_4}d the
EVPA is similar to that in Figure~\ref{fig:model_4}b at early and late
times, but is continuously CCW for $-17<t<+17$.  The total EVPA rotation
in Figure~\ref{fig:model_4}d, from g to h, is 199\deg.

%%%%%%%%%%%%%%%%%%%%%%%%%%%%%%%%%%%%%%%%%%%%%%%%%%%%%%%

%% FIGURE 9
\begin{figure}
\centering
\includegraphics[width=0.5\textwidth,trim=0cm 0.6cm 0cm 0.5cm]{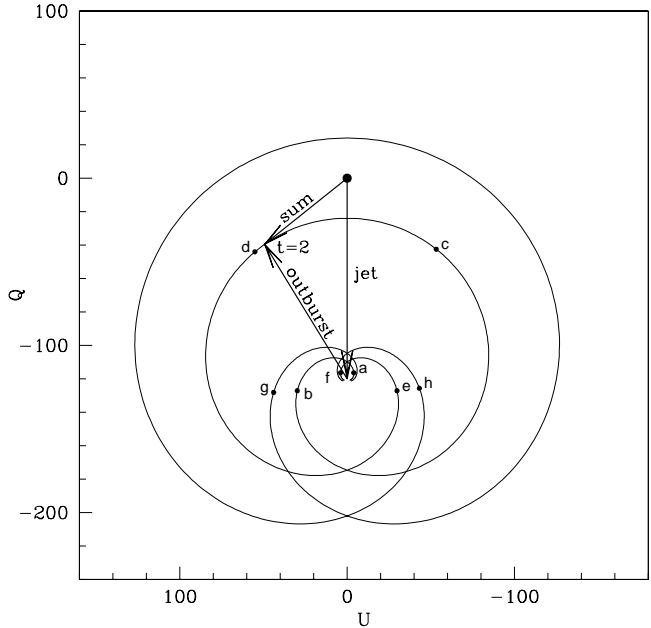}
\caption
{Stokes vectors corresponding to Figures~\ref{fig:model_4}b and
\ref{fig:model_4}d.  Vectors are drawn with amplitude $PF$ at
angle $\xi = \tan^{-1}(U/Q) = 2\times \mathrm{EVPA}$.  The vertical
vector is for the jet component in Figures~\ref{fig:model_4}a and
\ref{fig:model_4}c.  Vectors to the point labeled $t=2$ show the outburst
in Figure~\ref{fig:model_4}a at time $t=2$, and the sum (jet+outburst)
seen in  Figure~\ref{fig:model_4}b.  As time advances the outburst vector
rotates CCW around the tip of the jet vector, and the loop shows the
sum, starting at $t=-32$ and progressing through a ... f to $t=+32$.
The labeled dots on the loops mark extrema (max or min) in the angle
$\xi$.
\label{fig:stokes}
}
\end{figure}

%%%%%%%%%%%%%%%%%%%%%%%%%%%%%%%%%%%%%%%%%%%%%%%%%%%%

These changes are most easily understood with the Stokes Parameters.
Figure~\ref{fig:stokes} shows the Stokes plane for
Figure~\ref{fig:model_4}.  Here we adopt the IAU recommendations for the
sign of the Stokes parameters \citep{HB96} with the Stokes plane overlaid
on the sky plane; Q increases to the North, U increases to the East,
and the Stokes angle $\xi = \tan^{-1}(U/Q) = 2\times \mathrm{EVPA}$.

In Figure~\ref{fig:stokes} the vertical arrow labeled ``jet"
represents the steady jet, which is the same in Figures~\ref{fig:model_4}a
and \ref{fig:model_4}c. 
The Stokes vectors representing the outbursts are added to the jet vector,
to form the sum vectors, as shown at time t=2 for the weaker outburst
in Figure~\ref{fig:model_4}a. As time advances, the EVPA of the outburst
rotates CCW, and the sum vector traces out the inner loop.
The loop is parametric in time, and the times a--f on the loop
are EVPA reversal points that can be seen in Figure~\ref{fig:model_4}b.
At the reversal points the vectors are tangent to the loop.  The total
excursion of $\xi$ (between points c and d) is 102.6\deg; the EVPA
excursion, seen in Figure~\ref{fig:model_4}b, is 51.3\deg.

When the peak of the outburst is stronger than the jet, as in
Figure~\ref{fig:model_4}c, the loop encloses the origin, as shown by the
outer loop in Figure~\ref{fig:stokes}. In this case the sum vector rotates
continuously CCW between points g and h. The full excursion of $\xi$ is
398\deg, and the corresponding EVPA rotation in Figure~\ref{fig:model_4}d
is 199\deg.  This striking difference in EVPA rotation, caused by the
relative size of the jet and the outburst, could be responsible for
the differences in EVPA behavior seen in Figure~\ref{evpa8488}.
{\rm The 4.8 GHz outbursts are weak in 1985 and 1986;
whereas, at 8.0 and 14.5 GHz they are strong. It might be that
the outbursts are stronger than the jet at 8.0 and 14.5 GHz and weaker
at 4.8 GHz; and so on the Stokes plane the 8.0 and 14.5 GHz loops would
enclose the origin but the 4.8 GHz loop would not. This would give large
EVPA rotations at 8.0 and 14.5 GHz, with a small rotation at 4.8 GHz,
as seen in Figure~\ref{evpa8488}a.}

\subsection{Double Rotation with a Reversal} 
\label{s:double_rotation}

%%%%%%%%%%%%%%%%%%%%%%%%%%%%%%%%%%%%%%%%%%%%%%%%%%%

%%FIGURE 10
\begin{figure}
\centering
\includegraphics[width=0.5\textwidth,trim=0.1cm 0.6cm 0.5cm 0.5cm]{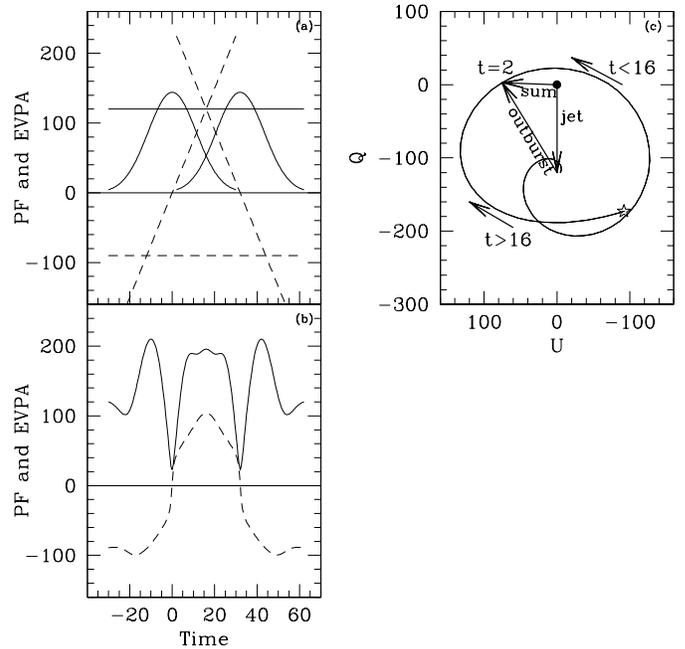}
\caption{
(a) Model using a steady jet and two outbursts, the first
with CCW EVPA rotation, and the second with CW rotation. The jet and
outbursts are the same as in Figure~\ref{fig:model_4}c except that
the EVPA of the second outburst has the opposite sense of rotation.
Solid lines are used for the polarized flux, $PF$, and dashed lines are
for the EVPA. 
(b) The result for adding together the three components in (a). 
(c) Stokes plane representation.  The jet vector is stationary. As time
advances, the outburst vector rotates CCW; its position at t=2 is 
shown. The sum vector, after $\rm t\approx -19$, rotates CCW to 
the star at t=16, where it reverses and follows its earlier path.
Note in (b) that the central part of the EVPA curve is much steeper
than the EVPA curves in (a). See text.
\label{fig:model_new3}
}
\end{figure}

%%%%%%%%%%%%%%%%%%%%%%%%%%%%%%%%%%%%%%%%%%%%%%%%%%%

A double rotation with a reversal can be obtained with two successive
outbursts, with opposite senses of rotation. Figure~\ref{fig:model_new3}
shows an example where both outbursts are stronger than the jet. In panel
(a) the two outbursts are the same as in Figure~\ref{fig:model_4}c
but with opposite rotations, and they are separated by $\Delta(t)
= 32$. The sum in (b) has some similaritites to Event D,
seen in Figure~\ref{evpa9404}. In both Event D and in the model
(Figure~\ref{fig:model_new3}b) the EVPA has rapid swings of order
180\deg, and $PF$ has a smooth top and deep minima centered near the
EVPA swings.

The Stokes plane plot for the model in Figure~\ref{fig:model_new3}b is in
Figure~\ref{fig:model_new3}c. The early part of this diagram is the same
as the corresponding part of Figure~\ref{fig:stokes}. When the second
outburst becomes appreciable, at $t\sim 2$, the loop opens out and,
at the star, where the second outburst begins to dominate the amplitude,
the loop reverses and goes CW back along the same track. This motion
gives the flat-top amplitude in Figure~\ref{fig:model_new3}b, and the
steep-sided EVPA curve.

Note that in Figure~\ref{fig:model_new3}b the central parts of the EVPA
swings are much steeper than the linear EVPA curves for the two outbursts.
In the context of models where the synchrotron source rotates around the
jet axis (Section~\ref{s:super}), this means that the physical rotation
rate can be much less than the apparent rate, seen as the rapid change
in EVPA.  It is likely that relativistic effects also affect
the apparent rotation \citep{B82}.

\subsection{Stokes Plot for Event D}
\label{s:stokes_eventD}

%%%%%%%%%%%%%%%%%%%%%%%%%%%%%%%%%%%%%%%%%%%%%%%%%%%%

%%FIGURE 11
\begin{figure*}
\centering
\includegraphics[width=0.9\textwidth,trim=0cm 9cm 0cm 0.5cm]{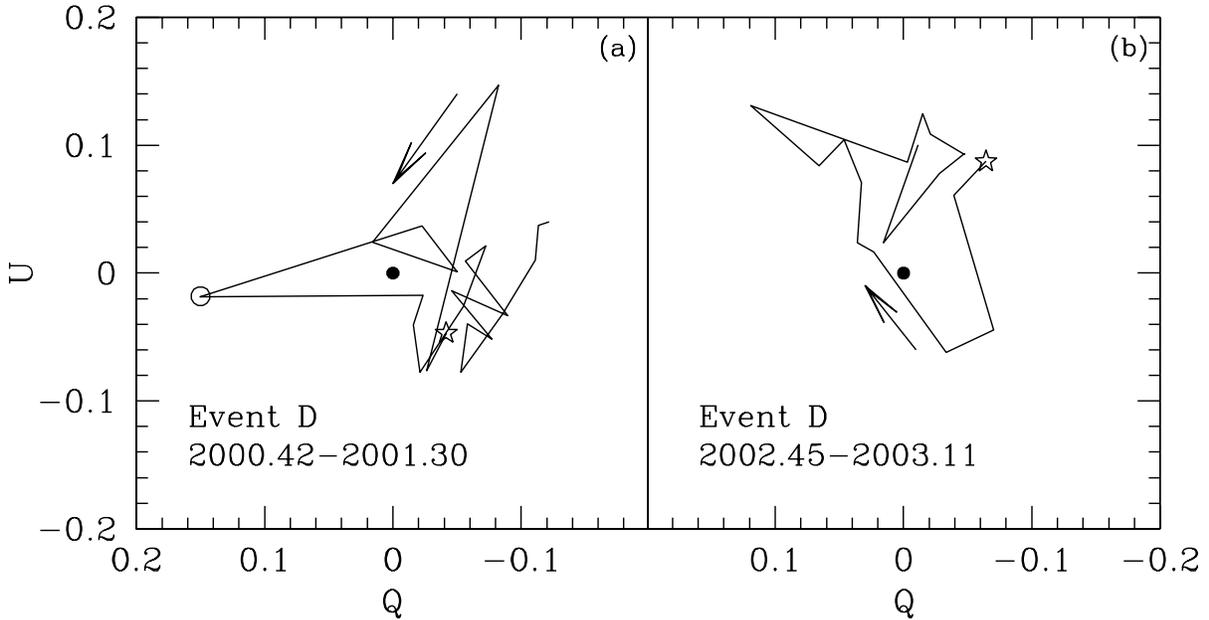}
\caption{Stokes plane representation for Event D at 14.5 GHz:
(a) the early, CCW, side of Event D; (b) the late, CW, side of event D.
The lines
start with the star and connect the tips of successive Stokes vectors,
which are not shown but point from the origin.  Time runs with the arrows,
and the total time interval is shown in the text.  Both loops enclose
the origin, so that the swing in $\xi$ is of order 360\deg, and the
EVPA rotation is of order 180\deg. See text.
\label{fig:stokes_eventD}
}
\end{figure*}

%%%%%%%%%%%%%%%%%%%%%%%%%%%%%%%%%%%%%%%%%%%%%%%%%%%%

In Figure~\ref{fig:stokes_eventD} we show the two sides of Event D
on the Stokes plane, for 14.5 GHz. Both of these loops are like the
outer loop in Figure~\ref{fig:stokes} in that they enclose
the origin. Hence the EVPA swing for each is of order 180\deg. In
Figure~\ref{fig:stokes_eventD}a the jumble of points near the star
contains both the beginning and the end of the swing. The circled
point is at 2000.94 and has $\xi=97\deg$ or $\mathrm{EVPA} = 48.5\deg$.
The polarization is exceptionally high for this point, but there is no
reason to exclude it as an outlier.  Because of it we can claim that
the loop surrounds the origin, and that the swing at 2001 is of order
180\deg.  On the CW side, shown in Figure~\ref{fig:stokes_eventD}b,
there are more points defining the loop, and, crucially, we see that in
Figure~\ref{evpa9404}a the swing is well-defined when data from all the
frequencies are included.

It is easier to study the EVPA with a time series like that in
Figure~\ref{evpa9404}a rather than with loops on the Stokes plane, because
time is not uniform on the loop. Further, it would be difficult to plot
all the frequencies together on the Stokes plane, because they would
need to be normalized in some way to the frequency spectrum of
the polarization. Still, the Stokes plot is useful in visualizing
how polarized radiations combine, and in arguing that, because the
loop encloses the origin, the EVPA rotation really is 180\deg~or more
\citep{Vil10}.

% GEO
%%%%%
\section{A Simple Geometry}
\label{s:geometry}

In the preceding Section we modeled the EVPA  rotation reversals with
a pair of outbursts whose EVPAs are counter-rotating.  We now present
a geometric model that can generate these counter-rotating outbursts,
in a simple and intuitive way.

Consider a plasma jet with a relativistic flow, with a helical magnetic
field.  Let a disturbance generate a sub-relativistic shock pair, a
forward shock traveling downstream and a reverse shock traveling upstream,
each with $\beta_\mathrm{sh}^\mathrm{jet}=0.1$ in the jet frame. Let
the Lorentz factor of the jet be $\Gamma_\mathrm{jet}^\mathrm{gal}
=10$ in the frame of the host galaxy.  Then in the galaxy
frame both shocks are moving forward relativistically, with
Lorentz factors $\Gamma_\mathrm{fwd}^\mathrm{gal}=11.05$ and
$\Gamma_\mathrm{rev}^\mathrm{gal} = 9.05$. An observer on axis sees
Doppler shifts of 22.1 and 18.1 for radiation from these shocks, and
if they have similar synchrotron sources then their flux density ratio
is about 1.5. The observed radiation from the reverse shock is not
substantially weaker than that from the forward shock.

Let the magnetic field lines have the structure of a right-hand
helix.  If the shocks travel along this helix then the forward shock
is seen to rotate CCW and the reverse shock (moving upstream in the jet
frame) is seen to rotate CW. With appropriate synchrotron sources whose
aspect to the axis is fixed, the EVPA rotation will follow that of the
shocks \citep{M08}.  In this geometry the observed rotation automatically
reverses when the second shock becomes the dominant source for the
polarized flux density. The right-handedness is required by the
observed sense of the EVPA reversal, CCW then CW.

% DELETED from first revision
%This tends to confirm the suggestion
%of a right-hand helix made by \citet{C17}, from an analysis of the
%evolution of the ridge lines of OJ\,287.}

In this model we have implicitly assumed that the plasma jet is not
rotating, so that the upstream shock is not carried into CCW rotation
as seen by the observer. But the jet is moving forward and cannot cross
the magnetic field. Hence, the helical field must be rotating at a rate
such that the screw action drives the plasma straight forward. The
non-relativistic condition for this is $\beta = \Omega R\cos \alpha$
where $\beta$ is the longitudinal jet velocity in units of $c$, $\Omega$
is the angular rotation rate of the magnetic field in $y^{-1}$, $R$
is the radius of curvature of the field in $ly$, and $\alpha$ is the
pitch angle.  But we have a relativistic flow and must be concerned
with the velocity-of-light cylinder around the axis. The situation here
is similar to that in a pulsar atmosphere, where the field lines bend
backward and the toroidal component of the field slips through the plasma
\citep{Mei12}. In this way the helical field continues across the light
cylinder while the plasma velocity stays below $c$.

We emphasize that we have proposed here a purely geometric model, and the
nature of the shock waves is not specified, nor is the mechanism by
which the source is guided by the helical field, and why the EVPA itself
stays fixed with respect to the helix.  In the next Section we describe
a physical model that has many of the features of the geometric model,
and suggest that it may explain the observations.

%NEW

%%%%%%%%%%%%%%%%%%%%%%%%%%%%%%%%%%%%%%%%%%%%%%%%%%%%%%5

%  from DLM 8/8/17 
%  converted from RTF to txt 
%\section{A Model Using a Super-Magnetosonic Jet} 

\section{Models Using Helically Magnetized Jets}
\label{s:super}
\subsection{Sub-fast, Super-slow Magnetosonic Jet Models}

\citet{NAK01} and \citet{NM04} simulated 1.5-D and 3-D helically
magnetized jets whose flow speed was slower than the jet's internal MHD
fast-mode magnetosonic wave speed, $V_{\rm jet} < V_{\rm fast} \approx
(V_{\rm A}^2 + V_{\rm s}^2)^{1/2}$, where $V_{\rm A}$ is the internal
jet Alfv\'en speed and $V_\mathrm{s}$ is the internal sound speed (with
$V_{\rm A} > V_{\rm s}$), but the jet was faster than the internal
slow-mode wave speed ($V_{\rm jet} > V_{\rm slow} \approx V_{\rm s}$).
Furthermore, the jet flow speed also was greater than the fast-mode
magnetosonic wave speed \textit{in the material into which the jet was
flowing}.\footnote{Often in numerical simulations of jets this material
represents the ``ambient medium".  However, in a jet with successive new
pulse or piston-like injections, the material in front of the contact
discontinuity is more likely to be \textit{prior} jet flow, which
in our model itself would have a helical magnetic field and a slower
flow speed.}

Jets like this, with a sub-magnetosonic internal Mach number but
super-magnetosonic external Mach number, develop three shocks in the
flow:  a forward fast-mode shock (FF), a forward slow-mode shock (FS),
and a reverse slow-mode shock (RS).  Furthermore, because of conservation
of the combined plasma and magnetic field angular momentum at
the FF shock, the material between the FF and FS shock has an enhanced
(compressed) helical magnetic field strength {\em and} a rotation velocity
significantly greater than the rotation rate of the main jet body near the
contact discontinuity and which also exceeds $V_{\rm slow}$.  Therefore,
if the supersonically-rotating plasma in the FS/FF region develops a
non-axisymmetric shock feature (e.g., near the FS shock itself), then
an observer viewing this jet end-on would observe synchrotron emission
from that feature that exhibited a {\em physical} rotation about the
line of sight of perhaps several radians.

Thus, a sub-fast, super-slow helically magnetized jet could be a promising
model for sources that exhibit a single, one-directional
rotation of the EVPA.  However, such jets do not produce
the double rotations, with reversals, seen in OJ\,287.

\subsection{Super-fast Magnetosonic Jet Models}

On the other hand, Nakamura et al. (2010) and Nakamura \& Meier (2014)
performed similar 1.5-D simulations of helically magnetized jets, but
whose flow speed this time was {\em greater} than the jet's internal
fast-mode magnetosonic wave speed.  These jets developed {\em four}
shocks in the flow: FF, FS, RS, and also a reverse fast shock (RF).
Figure 3d of Nakamura et al. (2010) shows that, in the galaxy frame, the
toroidal component of magnetic field is substantially enhanced between the
FF and FS shocks, and also between the RS and RF shocks.  This leads to
two moving synchrotron sources. Further, Figure 3e of this paper shows
that an azimuthal motion of the plasma is established between the FF
and FS shocks, and between the RS and RF shocks, but that the sense of
rotation is opposite in the two regions.  Thus, two oppositely--rotating
synchrotron--emitting regions are established, moving relativistically
downstream because $V_\mathrm{jet} > V_\mathrm{fast}$.  If the emission
regions are not axisymmetric, then an observer near the axis will see
the EVPAs of the two outbursts rotate in opposite directions.

This super-magnetosonic jet model is a good candidate to explain the
observations of the EVPA reversals seen in OJ\,287.  It produces the
main feature used in constructing Figure~\ref{fig:model_new3}, namely,
the two oppositely-rotating emission regions moving downstream.

\section{Optical Observations}
\label{s:optical}

Several research groups \citep{KIM88, DA09, Vil10, BP15} have reported
optical and IR observations of OJ\,287 that are closely-enough spaced in
time to be useful for studying EVPA rotations.  Two of them \citep{KIM88,
DA09} also show that the optical and radio variations of polarization are
nearly synchronous.  The observations of \citet{KIM88} occured at the same
time as our Event A, and have already been used in that discussion; see
Section~\ref{s:eventA}, and  Figures~\ref{fig:eventA} and \ref{evpa8488}.
To reduce confusion,
in the following we use nomenclature like `Figure~V1' to refer to Figure~1
in \citet{Vil10}, and `Figure~D2' to refer to Figure~2 in \citet{DA09}.

%%%%%%%%%%%%%%%%%%%%%%%%%%%%%%%%%%%%%%%%%%%%%%

%%FIGURE 12
\begin{figure}
\centering
\includegraphics[width=0.5\textwidth,trim=0cm 0.6cm 0cm 0.5cm]{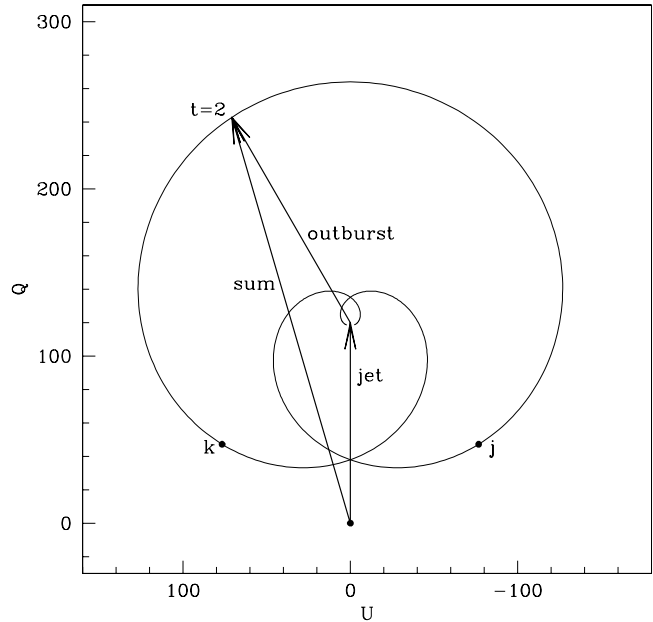}
\caption{As in Figure~\ref{fig:stokes} (outer loop) with the vector for the
jet rotated by 180\deg, which corresponds to an EVPA
rotation of 90\deg. The total swing of the resultant Stokes vector,
from j to k, is 116.7\deg, corresponding to an EVPA swing of 58.3\deg.
\label{fig:stokesrev}
}
\end{figure}

%%%%%%%%%%%%%%%%%%%%%%%%%%%%%%%%%%%%%%%%%%%%%%

Figure~V1 shows the R band flux density, the polarized flux density
$PF$, and the EVPA for OJ\,287, during 2004.9--2009.5. The EVPA has
points restricted to 0\deg -- 180\deg, and the figure is analogous to
our Figure~\ref{evpaorig}. It shows a number of rotations, several of
which are described in detail and are shown with Stokes vector plots
like the one in Figure~\ref{fig:stokes_eventD} for Event D.  The event
in April 2006 is shown expanded in Figure V2 and the Stokes plot is in
Figure V16.\footnote{\citet{Vil10} use axes with Q increasing from left to
right, opposite to our convention. This reverses the rotation direction on
the Stokes plane.} The loop in Figure V16 does not enclose the origin,
and the EVPA swing as seen in Figure V2 appears to be about 50\deg.
A better view of this event is in Figure D5, where the swing is seen to
be $\approx 45\deg$ CW. In this Figure the peak of $PF$ is in the middle
of the EVPA swing. This is the opposite of what happens in Event D,
seen in Figure~\ref{evpa9404}, where $PF$ has minima in the middle of
the EVPA swings.  An easy way to accommodate this difference, and keep
the optical result within the two-component model, is to shift the phase
of the EVPA.  Figure~\ref{fig:stokesrev} shows the Stokes plane for the
model in Figures~\ref{fig:model_4}c and \ref{fig:stokes} when the EVPA
of the jet is rotated by 90\deg; i.e. the Stokes vector for the jet is
reversed. Although the outburst is stronger than the jet, the
loop does not enclose the origin, and the CCW swing of the sum, from
j to k, is 116.7\deg.
Figure~\ref{fig:stokesrev} is analogous to Figure V16. Both
have the maximum $PF$ in the middle of the EVPA swing. However, any
physical significance attached to the shift of the jet EVPA relative to
the outburst will depend on the specific model used to describe the event.

From their Stokes plane plots like that in V16, \citet{Vil10} suggest that
OJ\,287 has two components of emission that generate the EVPA rotation,
the ``optically polarized core" (OPC) and the ``chaotic jet emmision".
Our model corresponds closely to this; the jet corresponds to the OPC,
and their chaotic jet emission corresponds to our outbursts that are
superposed on the jet. When, as in Figures V16 and V21, they show a
loop on the Stokes plane, it is their chaotic jet emission that has a
systematic CCW swing in EVPA. In Figure V16 the OPC is roughly lined up
with the maximum $PF$ as, in Figure~\ref{fig:stokesrev}, the vector for
the jet is aligned with the maximum $PF$.  This relationship also
holds for Figure V21 and the other Stokes plane plots in \citet{Vil10}.
Thus, we see that the two-component model that we use for the radio
observations may be useful for the optical data also.

\citet{Vil10} describe rotations in the EVPA of OJ\,287 at optical
wavelengths, but these are all single rotations, not double with a
reversal like those we have seen at radio wavelengths. However,
the EVPA data in Figure V1 are restricted to 0\deg~to 180\deg, and a
double rotation of order 180\deg~might not be recognized if it did exist
there. It would be useful to smooth the data in Figure V1 by adding $\pm
n\pi$ as needed, to search for further examples of rotation reversals
in OJ\,287.

In the literature, there are several examples of a large EVPA double
rotation with a reversal, at optical wavelengths.  The RoboPol program
\citep{Pav14} makes polarization observations of a large number of
AGN, with observations typically three days apart. Their plots show
two objects with double rotations with a reversal, J1806+694 (3C\,371)
\citep{BP15} and J1512-0905 (PKS 1510-089) \citep{BP16}.  J1512-0905
was also studied at R-band by \citet{BDA17}. There was no overlap
in these observations, but the Blinov et al observations ended with a
strong CW rotation and the Beakilini et al observations started about
20 days later with a strong CCW rotation. This appears to show a double
rotation of about 200 \deg, with a reversal. Figure 10 of \citet{BDA17}
shows a plot of the EVPA of J1512-0905 that combines the data from a
number of observers.

\section{Discussion}
\label{s:discussion}

\subsection{Time Scales for EVPA Rotation}
\label{s:time_scales}

We have found various rotation rates in OJ\,287, from the fastest,
$17\deg\,d^{-1}$ in the CW swing in Event~A (Figure~\ref{fig:eventA}),
to the over-all long-term trend of roughly 90\deg~in $30~y$
(Figure~\ref{evpa7416}).  The reciprocal of a rate is a time scale,
which we take here to be the time to rotate by one radian.  Thus our time
scales run from 3.3 days to 30 years. On the short end, measureable time
scales are limited by the sampling interval, which for \citet{KIM88}
is one day. \citet{LBPP17} have argued that for reliable recovery of
the intrinsic time scale, the sampling interval should not exceed $\sim
30\%$ of the intrinsic scale; i.e. we need at least three samples per
intrinsic time scale.
Thus $17\deg\,d^{-1}$ is about the fastest
rate that \citet{KIM88} could have reliably determined. The UMRAO
points are typically 3 days apart and so $5\deg\,d^{-1}$ is about the
fastest that can be found in the UMRAO data.  The MOJAVE points are a
few weeks to several months apart and the most rapid CCW and CW swings
in the rotation reversal events cannot be determined reliably from the
MOJAVE data alone.  This has already been noted in Section~\ref{s:obs}.
The CCW rate in Event D is about $1.8\deg\,d^{-1}$, and its detection
requires a sampling interval of 10 days or less.

To find a time scale in the frame of the jet, $\tau_\mathrm{jet}$, we
multiply the observed time scale $\tau_\mathrm{o}$ by $\delta/(1+z)$,
where $\delta$ is the Doppler factor of the radio source. There are two
different values for $\delta$ in the literature, and we refer to them as
``early" and ``recent". Two early values are $\delta=17.0$ \citep{Hov09}
and $\delta=18.9\pm 6$ \citep{JM05}, and they are derived from 43-GHz
flares in 2003 and 1998-2000, respectively. Two recent measurements
both give $\delta=8.7$ \citep{JM17,LMA17}, and they are derived from
mm-wave flares after 2007.  Apparently, $\delta$ changed around 2005;
why? We suggest that there was a change in the direction of the
inner jet. \citet{AMJ12} showed that the PA of the inner jet jumped
in 2004 at 43 GHz; and at 15 GHz there was a similar change in 2006
\citep{C17}. Presumably, this PA jump reflects an increase in the viewing
angle to the jet, and as a consequence the Doppler factor changed by a
factor of about 2.  Since all the major rotation/reversal events at 15
GHz that we see in Figure~\ref{evpa7416} happened prior to 2006, we use
the early value, $\delta\approx 17$ in the following discussion. However,
the use of the lower value would not make a substantive change.

With $\delta\approx 17$, $\tau_\mathrm{jet}/\tau_\mathrm{o} \approx
13$. The fastest swing, $17\deg\,d^{-1}$ now becomes the shortest time
scale in the jet, $\tau_\mathrm{jet,min} \approx 44~d$.
This time scale may be too short to represent a shock circulating
around a helix, as it would make the radius of curvature of the field
line much less than $1~ly$.  Our preferred explanation, using the model
in Section~\ref{s:stokes_vectors}, has no analogous velocity-of-light
limit. As seen in Figure~\ref{fig:stokes}, the rotation speed of the
resultant Stokes vector can be very high, when the amplitudes of
the two components are nearly the same.

Longer timescales are seen in the slowly-changing EVPA baseline in
Figure~\ref{evpa7416}a; e.g., in 1993 where the apparent rate is
about $20\deg~y^{-1}$, or $\tau_{jet}\sim 50~y$.
The changes in 2005 -- 2012 are coincident with changes in the orientation
of the inner jet, as defined by the appearance of a new superluminal
component \citep{C17}.

\subsection{Outbursts Without Rotations}

In Figure~\ref{evpa8488} Event A appears to be associated with
the strong outbursts A1 and A2 in flux density, but when we look
at Figure~\ref{evpa7416} we see outbursts in 1981--1985 that are not
associated with an EVPA rotation. Similarly, a series of modest outbursts
in 2004 -- 2009, and larger ones in 2009 -- 2012, are not associated with
large rotations. This may reflect a selection effect. In the two-component
Gaussian model, a large rapid rotation is only seen when the conditions
are right; the outburst must be stronger than the jet, and the
EVPA phase and rotation rate must be appropriate. On the other
hand, outbursts without large rotations may simply show that there is
more than one cause for the outbursts.

\subsection{Optical Flares with a 12-year Period}

The rotation reversal events A, B, C, and D occur at roughly 1986.1,
1990.0, 1998.6, and 2001.8, respectively. The intervals A--C and B--D
are both roughly 12 years.  This is interesting, because the optical
flares that match a binary black-hole model have a period of about 12
years \citep{Val11}. The top axis of Figure~\ref{evpa7416} has six
bars that indicate the epochs of the flares. The epochs are 1983.0,
1984.2, 1994.8, 1996.0, 2005.8, and 2007.7 \citep{Sill88, Val06,
Sill96a, Sill96b, Val08a, Val08b}.  These are the original references
reporting the flares, except for 1984.2, where the reference is to the
compilation in \citet{Val06}.  Additionally, a strong flare was seen at
2015.9 after having been predicted \citep{Val16}; showing that the model
closely matches the observations.  We also note that light
curves for OJ\,287 are highly variable, and that an analysis of 9.2
years of well-sampled optical data yielded evidence for quasi-periodic
oscillations of periods $\sim 400$ and $\sim 800$~days \citep{BZS17}.

In Figure~\ref{evpa7416} the optical pairs in 1983-84 and 1994-96 precede
the radio rotation pairs A-B and C-D by about three years. However, there
are no radio events corresponding to the optical pair in 2005-07, and this
suggests that the radio--optical 12--year similarity is a coincidence.

\section{Summary and Conclusions}
\label{s:conclusions}

We report what appears to be a new phenomenon, rotation reversals of
the EVPA at radio frequencies, in the BL Lac object OJ\,287. These
consist of a large $\sim$180\deg~CCW rotation followed by a similar CW
rotation. Three of these events were seen in 40 years, and a fourth,
smaller one, was also seen.  They were all in the same direction, CCW
followed by CW.  We suggest that a rotation can be explained with a
two-component model consisting of an outburst superposed on a steady jet,
with the EVPA of the outburst rotating steadily in time.  This reproduces
many of the observed features of a rotation.  A three-component model
consisting of two successive outbursts with oppositely rotating EVPAs,
together with the jet component, explains the reversals.  This model is
also applicable to polarization rotations seen at optical wavelengths.

In a more physical model, we consider that the reversals take place
in a super-magnetosonic jet; i.e., one in which the bulk speed of
the plasma is greater than the speed of the fast magnetosonic wave.
The jet is threaded by a helical magnetic field.  We use the mechanism
analyzed by \citet{NGM10} and \citet{NM14} that produces four MHD waves,
forward and reverse fast and slow magnetosonic waves.  Between the
forward fast and slow waves the toroidal component of magnetic field
is compressed; this increases the angular momentum of the field, and
to conserve angular momentum the plasma rotates around the axis in the
opposite direction. This happens also to the reverse fast and slow pair of
magnetosonic waves, but the rotation is in the opposite sense. This forms
two regions of enhanced plasma density and magnetic field, rotating in
opposite directions. Both regions move relativistically downstream because
$V_{\rm jet} > V_{\rm fast}$.  The resulting synchrotron radiation,
as seen by an observer near the axis, consists of two outbursts that
have oppositely rotating EVPAs.
The observed rotation sense, CCW followed by CW, requires a right-hand
helix.

We conclude that our observations of EVPA reversals provide evidence
for a strong helical magnetic field in OJ\,287.  This is consistent with
the observations and conclusions of many others e.g.~\citet{C15,
GLB16,MG17}. The observations also provide evidence that the jet of OJ\,287
is super-magnetosonic, and this can provide a constraint on $B^2/n$,
where $B$ is the strength of the magnetic field and n is the particle
density.

%%%%%%%%%%%%%%%%%%%%%%%%%%%%%%%%%%%%%%%%%%%%%%%%%%%%

\acknowledgments
We thank the referee, D.C. Gabuzda, for a thorough reading of the
manuscript and for many suggestions that improved the paper.  We are
grateful to D.E. Homan and S. Kiehlmann for reading the manuscript and
making useful suggestions; and to the entire MOJAVE team for their years
of work in assembling the data used here.  This research has made use
of data from the MOJAVE database that is maintained by the MOJAVE team
\citep{Lis09}; the MOJAVE program is supported under NASA-Fermi grant
NNX15AU76G .  Operations at UMRAO were supported by the University of
Michigan, and research there was funded in part by NASA-Fermi GI
grants  NNX09AU16G,  NNX10AP16G, NNX11AO13G, NNX13AP18G and NSF grant
AST- 0607523.  YYK and ABP were supported by the Russian Foundation for
Basic Research (project 17-02-00197), the Basic Research Program P-28
of the Presidium of the Russian Academy of Sciences and the government
of the Russian Federation (agreement 05.Y09.21.0018).  TS was funded by
the Academy of Finland projects 274477 and 284495 This research has been
supported in part by the Alexander von Humboldt Foundation, and it has
made use of the Swinburne University of Technology software correlator,
developed as part of the Australian Major National Research Facilities
Programme and operated under license \citep{Del07}.

%%%%%%%%%%%%%%%%%%%%%%%%%%%%%%%%%%%%%%%%%%%%%%%%%%%%

%REF

\end{document}